\documentclass{article}
\usepackage{natbib}
\usepackage{graphicx}
\usepackage[english]{babel}
\usepackage[latin1]{inputenc}
\usepackage{amssymb,amsmath}
\usepackage{amsfonts}
\usepackage{booktabs}
\usepackage{url}
\usepackage{geometry}
\title{Multistability and rare spontaneous transitions in barotropic $\beta$-plane turbulence}

    \author{
    Eric Simonnet\footnote{INPHYNI, UMR7010 CNRS-UNS, 1361, route des Lucioles, 06560 Valbonne.    
\thanks{eric.simonnet@inphyni.cnrs.fr}},
     \\
    Joran Rolland\footnote{Current address: Univ. Lille, Centrale Lille, UMR CNRS 9014 - LMFL - Laboratoire de M\'ecanique des Fluides de Lille - Kamp\'e de F\'eriet, France.} and Freddy Bouchet
\footnote{Univ Lyon, Ens de Lyon, Univ Claude Bernard, CNRS, Laboratoire de Physique, Lyon, France}}

\begin{document}

\maketitle

\begin{abstract}
We demonstrate that turbulent zonal jets, analogous to Jovian ones, which are quasi-stationary, are actually metastable. After extremely long times, they randomly switch to new configurations with a different number of jets. The genericity of this phenomenon suggests that most quasi-stationary turbulent planetary atmospheres might have many climates and attractors for fixed values of the external forcing parameters. A key message is that this situation will usually not be detected by simply running the numerical models, because of the extremely long mean transition time to change from one climate to another. In order to study such phenomena, we need to use specific tools: rare event algorithms and large deviation theory. With these tools, we make a full statistical mechanics study of a classical barotropic beta-plane quasigeostrophic model. It exhibits robust bimodality with abrupt transitions. We show that new jets spontaneously nucleate from westward jets. The numerically computed mean transition time is consistent with an Arrhenius law showing an exponential decrease of the probability as the Ekman dissipation decreases. This phenomenology is controlled by rare noise-driven paths called {\it instantons}. Moreover, we compute the saddles of the corresponding effective dynamics. For the dynamics of states with three alternating jets, we uncover an unexpectedly rich dynamics governed by the symmetric group ${\cal S}_3$ of permutations, with two distinct families of instantons, which is a surprise for a system where everything seemed stationary in the hundreds of previous simulations of this model. We discuss the future generalization of our approach to more realistic models.
\end{abstract}

%
\section{Introduction}

Many complex physical and natural systems have more than one local attractor.
The dynamics can then settle close to one of these local attractors for
a very long time, and randomly switch to another one. Well-known
examples are phase transitions in condensed matter or conformational
change of molecules in biochemistry \citep{chem}. Such situations
of multistability are ubiquitous also in geophysics and for turbulent
flows \citep{turb,Bouchet_Simonnet_2008}. The reversal of the Earth
magnetic field on geological timescales, which is due to turbulent
motion of the Earth's metal core \citep{dynamo}, is a paradigmatic
example.

The climate system is no exception and global multistability and abrupt
transitions existed from the global Neoproterozoic glaciations (snowball
Earth events) \citep{pierrehumbert2011climate} to glacial-interglacial
cycles of the Pleistocene \citep{paillard1998timing}. Very often,
those transitions are due to the internal dynamics and are not caused
by changes of external parameters. This is the case for instance for
the fast Dansgaard--Oeschger events \citep{Dansgaard1993,ditlevsen2007climate,DO}.
More local multistabilities also exist in part of the climate system,
for instance the Kuroshio bimodality \citep{Ku1,Ku2} in the North
Pacific.

If multistability and abrupt transitions between distinguishable states
are generic features of complex dynamical systems, why should planetary
atmosphere dynamics be an exception? Actually Lorenz was already raising
this question in 1967 \citep{Lorenz1967}. The hypothesis of a transition
to a superrotating atmosphere with eastward equatorial jets has been
addressed many times \citep{Held1999} and observed in many numerical
experiments \citep{arnold2012abrupt}. Recently the more specific question
of an abrupt transition to superrotation has been studied carefully
\citep{herbert2020atmospheric}. On Earth, superrotation may have played
a role in the climate of the past: it was observed in numerical simulations
of warm climates such as the Eocene \citep{Caballero2010}, and it
has been suggested that it could explain the permanent El Ni\~no conditions
indicated by paleoclimatic proxies during the Pliocene \citep{Tziperman2009}.

The aim of this paper is to address the hypothesis of planetary atmosphere
multistability and abrupt transitions not for equatorial jets but
for midlatitude eddy-driven jets. This is a natural hypothesis as
multiple midlatitude jets are observed on many planets, and that even
with Earth conditions slight changes of parameters lead to a different
number of jets \citep{Lee,lee2005baroclinic} and jets of different
nature \citep{Kaspi}. Because it is difficult to change from a state
with $n$ zonal jets, to a state with $n+1$ zonal jets continuously,
it is natural to expect multistability, abrupt transitions, and spontaneous
transitions between those different states to be generic.

\begin{figure}[htpb]
\centerline{ \includegraphics[width=12cm]{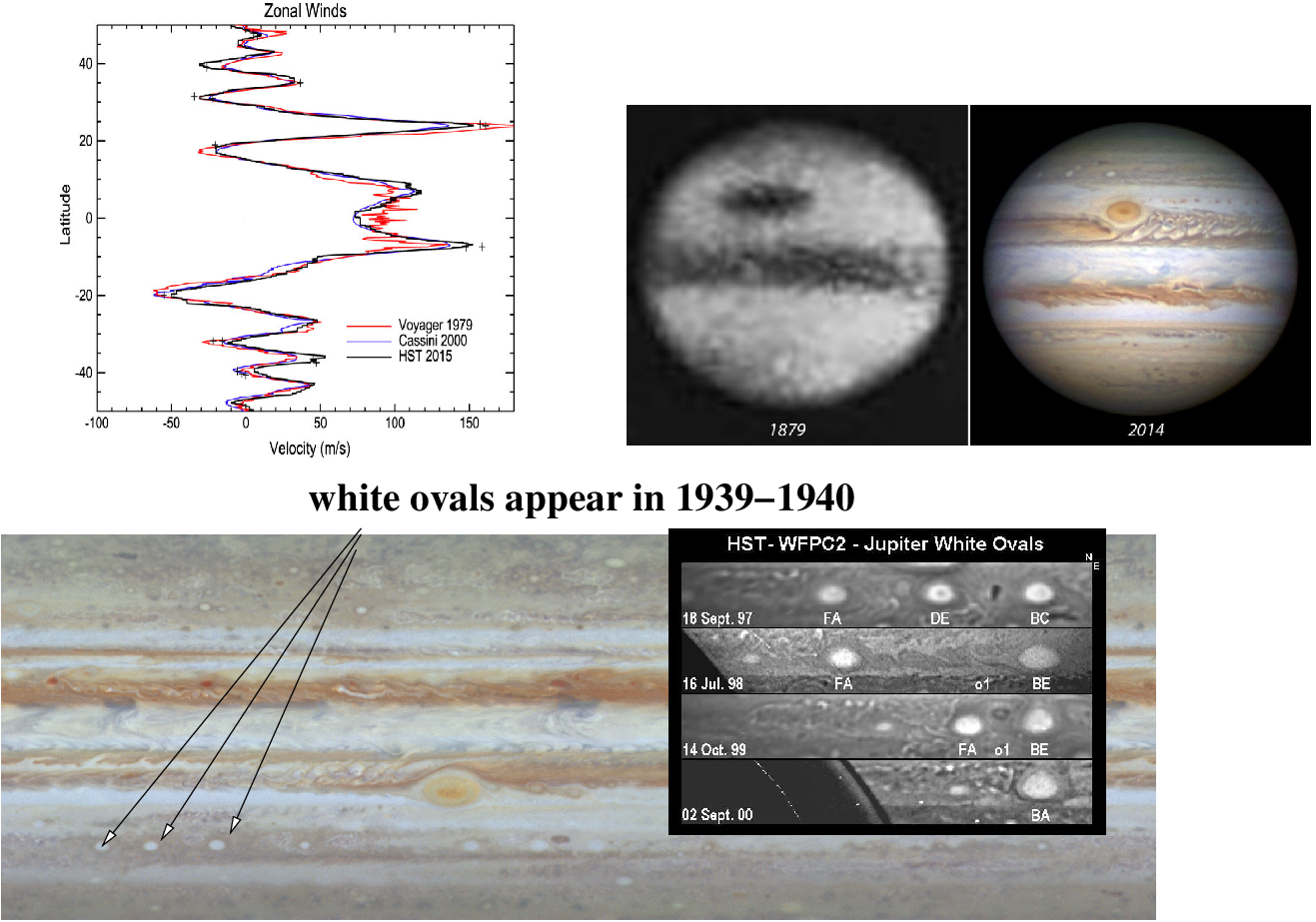}}
\caption{Upper left panel: Jupiter's zonally averaged zonal velocity profile observed by Voyager, Cassini \citep{Porco} and Hubble \citep{simonetal}. Upper right panel: Great Red Spot contraction after one century (left: from \citep{clerke}, right: NASA). Lower panel: three white ovals appeared in 1939-1940
\citep{Rogers} and later merged to from the oval BA, also called "Red Spot Junior" in 2006 when it turned red.}
\label{grsobs}
\end{figure}

A striking aspect of Jovian jets is their stability. Despite the strong turbulent activity
at small scales, one hardly notices any difference between the two
separate measurements by Voyager and Cassini 20 years apart (see Fig.\ref{grsobs}).
The Jovian large-scale atmosphere, including the Great Red Spot,
in fact appears to be stationary for decades. However, a fantastic
event occurred in 1939-1940 when Jupiter lost one of its jets \citep{Rogers}.
Three white anticyclones were created which started to merge in the
late 90s into a large white anticyclone (see Fig.\ref{grsobs}). As
such, \cite{Markus} called the 1939--1940 event a global climate
change on Jupiter. The mechanism responsible for the disappearance
of one of the jet remains completely unknown.\\

A key scientific aspect of bistability situations is that the time
scale for observing a spontaneous transition from one attractor to
another can be extremely long. For instance on Jupiter, no such events
as the 1939-1940 jet disappearance occurred since then. The typical
timescale for observing such transitions is thus longer than 80 years, while
energy balance for jet-mean flow interactions, Rossby wave dynamics,
and radiative effects have time scales much smaller ranging from one
Jupiter day to several years. For this reason, we argue in this paper
that most of time, scientists running numerical simulations are not
aware that their model may have more than one attractor, just because
they never simulate it long enough. Actually, simulating the model
long enough and waiting, is most of the time impossible with currently
available computational capabilities. This time-scale separation issue
seems an impassable barrier. Breaking this constraint and studying
unobserved phenomenologies for such transitions are two fascinating
scientific aims. For instance in this paper, we will demonstrate that
a barotropic model of self organized jets has a huge number of climate (attractors)
for a fixed set of boundary conditions and forcing parameters, with
an amazingly rich long-time dynamical structure for the transitions
from one attractor to another. This was completely unexpected and
never observed, even if analogous models have been run by many investigators.

In order to face this time scale separation issue, we need to build
completely new theoretical and numerical tools. For this purpose, it is scientifically
sound to go back to basics and to start with the simplest possible
model that contains a phenomenology analogous to Jovian troposphere dynamics.
We will consider forced--dissipative barotropic turbulence on a beta
plane, in which small-scale turbulence self-organizes into large-scale
coherent structures, with upscale fluxes of energy. Although such
a barotropic description of Jovian troposphere has long been used for theoretical studies \citep{galperin2014cassini}, it has clearly some strong limitations
as the energy transport through baroclinic instability is described only roughly by the stochastic forcing, as models with baroclinic
instability may lead to dynamics with different qualitative aspects
\citep{jougla2017energetics}, and as the barotropic model does not contain vorticity stretching terms. It might be interesting to consider such a barotropic model for the evolution and stability of Jovian's interior jets, as first analyzed by \cite{ingersoll1982motion} and more recently by \cite{Kaspi}. In any case, our aim in this paper is not to model precisely Jupiter. This work is rather a relevant study of the multistability of the barotropic model.
As far as multistability for Jovian planets are concerned, this work must be understood as a very first step in developing the requested methodology to study multistability with more realistic models.

Statistical approaches such as S3T actually allowed to
predict the non-uniqueness of solutions (see \cite{FI1}, or \cite{parker2013zonal} for the transition
regime to zonal jets). Moreover, wonderful zonal jet experiments have
demonstrated bistability in fairly barotropic regimes \citep{lemasquerier2020zonal}.
However, no spontaneous transitions between attractors were observed
before the work of \cite{Bouchet_Rolland_Simonnet_2019:C}. In this paper,
with the beta-plane barotropic model with stochastic forcing, we were
able to observe multistability between states with two eddy-driven
zonal jets and states with three eddy-driven zonal jets, and to observe
spontaneous transitions between those two states. Because of the huge
numerical cost, it would have been impossible to observe several of
these very rare transitions, and to study their dynamics and probability.
In order to face this critical practical problem, we have to use a
rare event algorithm in order to concentrate the computational cost
on transitions trajectories from one attractor to another, rather
than on extremely long period where nothing dynamically interesting
occurs. Rare event algorithms aim at this goal \citep{KH,PdM}. In
a dynamical context, they have been first applied for complex and
bio-molecules, see for instance \cite{metzner2009transition,hartmann_characterization_2013}.
More recently, in order to progressively go towards genuine geophysical
applications, they have been applied to Lorenz models \citep{WOUTERS:2016:A},
partial differential equations \citep{Joran}, turbulence problems
\citep{Grafke,Laurie,Grauer,Bouchet_Rolland_Simonnet_2019:C,lestang2020numerical},
geophysical fluid dynamics \citep{Bouchet_Rolland_Simonnet_2019:C},
and climate applications \citep{RAGONE:2018:A,webber_practical_2019,ragone2019computation,plotkin2019maximizing}.
Some approaches through minimum action methods, related to large deviation
theory, are reviewed in \cite{grafke2019numerical}. In \cite{Bouchet_Rolland_Simonnet_2019:C},
and in this paper, we rather use the Adaptive Multilevel Splitting
(AMS) algorithm, a very simple rare event algorithm which is well
suited to study rare transitions \citep{CerouGuya} (see \cite{cerou_adaptive_2019}
for a short review, and \cite{Joran,Bouchet_Rolland_Simonnet_2019:C}
for application for partial differential equations and geostrophic
turbulence problems). In \cite{Bouchet_Rolland_Simonnet_2019:C},
thanks to the AMS algorithm, we were able to demonstrate that rare
transition paths concentrate close to instantons and that transition
rates follow an Arrhenius law. The concentration of transition paths
close to predictable trajectories, called instantons, is a fascinating
property shared by many rare events. This has been now been observed
in several turbulent flow applications \citep{Laurie,Grafke,BLZ,dematteis2019experimental}
and other geophysical applications, for instance the dynamics of the
solar system \citep{woillez2020instantons}. The idea that Jovian's
jet transitions should follow instantons and be described by Arrhenius
laws, like condensed matter phase transitions, is really striking.
\\

The aim of this paper is to develop and extend those results in several
ways, pushing the power of rare event algorithm for studying unobserved
phenomena so far, for instance rare transitions for eddy-driven jets.
We clearly demonstrate the generic nature of multistability by performing
bistability experiments between the attractors with two and three
jets respectively. New results include a complete description of the
three-jet structure. Amazingly, there are actually six different types
of three-jet states with different jet spacing. The dynamics can remain
quasi-stationary close to the two-jet state for a very long time,
switch to the three-jet states; change several times its type of three-jet
states, before coming back to the two-jet states. We confirm the instanton
phenomenology for transition between the two-jet states and the three
jet-states, but we also demonstrate it for transitions between different
three-jet states. This fascinating complex phase space structure for
the slow dynamics, occurring on very long time scales, has profound
implication on our understanding of gas-giant planets. It suggests
that the current state of Jovian troposphere, and of many other gas-giant planets, is most probably one
among many different ones, for a fixed set of forcing parameters.
This profound fact suggests a very simple explanation of the observed asymmetry
between the northern and southern hemisphere structure of Jupiter.
In the conclusion, we explain why we believe that this observation
for a barotropic dynamics is expected to be robust for more comprehensive
models. This opens a completely new set of scientific questions for
planetary atmosphere studies.

Another new result of this paper concerns the hypothesis of barotropic
adjustment of eddy-driven jets. An eddy-driven jet with a time-scale
separation between eddy dynamics and zonal jet time scale, should
be unstable in order to transfer meridionally zonally-averaged momentum,
while it should be at the same time stable if quasi-stationary zonal jet
states are observed for very long times. This apparent contradiction
leads to the hypothesis of barotropic adjustment: the state of the
system should be marginally stable (or unstable) in order to fulfill
these two seemingly contradictory requirements. The relevance of this
adjustment hypothesis has been recently discussed in the context of
Jovian planets and for a hierarchy of models \citep{read2020turbulent,read2020baroclinic}.
Through an empirical analysis of the Rayleigh-Kuo criterion, we demonstrate
in this paper that the zonal jets actual constantly remain in a state
of marginal stability. The striking new results is that this remains
true also during the transitions between different attractors: The
nucleation of a new jet, or the merging of two jets, both occur within
barotropically adjusted states.

The paper is organized as follow. We describe the barotropic quasigeostrophic model and its
phenomenology in Section 2. Section \ref{sams}\ref{sbim} show some bimodality results
involving transitions between two-jet and three-jet solutions. Section \ref{sams}\ref{samsb} study in more
details these rare transitions using an advanced rare event algorithm (AMS). Section \ref{sams}\ref{samsc} discusses the
barotropic adjument and Rayleigh-Kuo criterion together with the marginal stability of the dynamics.
Section \ref{sams}\ref{samsd} focuses on the saddles of the effective dynamics.
Section~\ref{internald}
focuses on the dynamics of the three-jet states only and a summary
of the full effective dynamics is given. Section~\ref{ldp} addresses
the question of the Arrhenius law when the Ekman dissipation $\alpha$
decreases. We then conclude in Section~\ref{sconclusion}.\\


\section{The beta-plane model for barotropic flows and zonal jet dynamics}

We consider in the following the barotropic quasi-geostrophic equations, with a beta plane approximation for the variation of the Coriolis parameter. The equations in a doubly periodic domain $\mathcal{D}=[0,2\pi L)\times[0,2\pi L)$ read
\begin{equation}
\partial_t \omega +\mathbf{v}\cdot\mathbf{\nabla}\omega + \beta_d v_{y} =-\lambda\omega-\nu_{n,d}\left(-\Delta\right)^{n}\omega+\sqrt{\sigma}\eta,\label{eq:barotropic-d}
\end{equation}
where $\mathbf{v}=\mathbf{e}_{z}\times\mathbf{\nabla}\psi$ is the non-divergent velocity, $v_{y}$ the meridional velocity component, $\omega=\Delta\psi$ and $\psi$ are the vorticity and the stream function, respectively. $\lambda$ is a linear friction coefficient, $\nu_{n,d}$ is
a (hyper-)viscosity coefficient, and $\beta_d$ is the mean gradient
of potential vorticity. $\eta$
is a white in time Gaussian random noise, with spatial correlations
\[
\mathbf{E}\left[\eta({\bf r}_{1},t_{1})\eta({\bf r}_{2},t_{2})\right]=C({\bf r}_{1}-{\bf r}_{2})\delta(t_{1}-t_{2})
\]
that parametrizes the curl of the forces (physically due, for example, to the effect of baroclinic instabilities or convection).
The correlation function $C$ is assumed to be normalised such that
$\sigma$ represents the average energy injection rate, so that the average
energy injection rate per unit of area (or equivalently per unit of mass taking into account density and the layer thickness) is $\epsilon=\sigma/4\pi^{2}L^{2}l_{x}$.

For atmospheric flows, viscosity is often negligible in the global energy balance and this is the regime that we will study in the following.
Then, the main energy dissipation mechanism is linear friction. The evolution of the average energy (averaged over the noise realisations) $E$ is given by
\[
\frac{dE}{dt}=-2\lambda E+\sigma.
\]
In a stationary state, we have $E=E_{stat}=\sigma/2\lambda$, expressing the balance between forces and dissipation.
This relation gives the typical
velocity associated with the coherent structure $U_r \sim\sqrt{E_{stat}}/L\sim\sqrt{\epsilon/2\lambda}$.
As will be clear in the following, we expect the non-zonal velocity perturbation to follow an inviscid relaxation, on a typical
time scale related to the inverse of the shear rate. Assuming that a typical vorticity or shear is of order $s= U_r/L$
corresponding to a time $\tau=L/U_r$, it is then natural to define a non-dimensional parameter $\alpha$ as the ratio
of the shear time scale over the dissipative time scale $1/\lambda$,
\[
\alpha=\lambda\tau=L\sqrt{\frac{2\lambda^{3}}{\epsilon}}.
\]
When $\beta$ is large enough, several zonal jets can develop in the domain. An
important scale is the so-called Rhines scale $R$ which gives the typical size of the meridional jet width:
\[L_{R}=\left(U_r/\beta_d \right)^{1/2}=\left(\epsilon/\beta_d^{2}\lambda\right)^{1/4}.\]

We write the non-dimensional barotropic equation using the box size $L$ as a length unit and the inverse of a typical
shear $\tau=L/U_r$ as a time unit. We thus obtain (with a slight abuse of notation, due to the fact that we use the same symbols for the non-dimensional fields):
\begin{equation}
 \partial_t \omega +\mathbf{v}\cdot\mathbf{\nabla}\omega + \beta v_{y} =-\alpha\omega-\nu_{n}\left(-\Delta\right)^{n}\omega+\sqrt{2\alpha}\eta,\label{eq:barotropic}
\end{equation}
where, in terms of the dimensional parameters, we have $\nu_n=\nu_{n,d}\tau/L^{2n}$, $\beta=\beta_d L \tau$. Observe that the above equation is defined on a domain $\mathcal{D}=[0,2\pi l_{x})\times[0,2\pi )$ and the averaged stationary energy for $\nu_n\ll \alpha$ is of order one. Please observe that the non dimensional number $\beta$ is equal to the square of the ratio of the domain size divided by the Rhines scale. As a consequence, according to the common belief, the number of jets should scale like $\beta^{1/2}$ when $\beta$ is changed.

The phenomenology of stochastically forced barotropic turbulence has been described for a very long time in the literature \citep{Vallis1,dritschel2008multiple,galperin2001universal,bakasioannou2013,tobias2013direct}. Let us remind briefly the main aspects. Starting from rest, as one forces randomly at scales smaller than the Rhines scale, a 2D inverse energy cascade develops during a transient stage. This cascade is stopped at a scale of the order of the Rhines scale, where Rossby waves start to play a fundamental role and energy transfer is inhibited. It is well-known that this barrier is anisotropic and take the form of a dumbbell shape which favor zonal structure \citep{Vallis1}. Zonal jets then form, with a typical width of order of the Rhines scale, while jet waves often called modified Rossby waves, or Zonons \citep{galperin2001universal}, play an important dynamical role.

\cite{galperin_read_2019} gives a survey of the latest theoretical, numerical and experimental
advancements atmosphere and ocean jets dynamics and their interactions
with turbulence. Once jets are formed, the cascade framework is not relevant anymore. As illustrated by many contribution,
the beta-plane barotropic model has been a test-bed for many phenomenological
and theoretical studies of jet dynamics, for instance, second-order
closures (also called S3T or CE2 expansions) \citep{farrellioannou2003,Srinivasan-Young-2011-JAS}
or analogies with pattern formation \citep{parker2013zonal}. Second-order closures, or the related quasilinear approach, has been
argued to be exact in the limit of time scale separation \citep{Bouchet_Nardini_Tangarife_2013_Kinetic_JStatPhys}
between the zonal jet time scale on one hand, and the eddy relaxation
time on the other hand. This is a limit relevant for Jupiter. Moreover, \cite{Woillez} has given a fairly complete analytical description
of the zonal jet structure of barotropic jets, including a dynamical explanation of the
westward-eastward jet asymmetry, and a precise description of the
westward jet cusps and their regularization.
For values of $\alpha$ of order one or larger, the system settles in a statistically stationary state with such a phenomenology. For smaller values of $\alpha$, the jets become very strong. Once such strong jets are formed, most of the times they progressively expel the waves; sometimes a few neutral mode can stay or appear.
For smaller values of $\alpha$ no wave exist anymore, and both the inverse cascade phenomenology and the modified Rossby wave phenomenology become irrelevant. The system then settles in a statistically stationary state, where the dynamics is dominated by the strong zonal jets, the average width of which is still approximately given by the Rhines scale, and with fluctuations of order $\sqrt{\alpha}$. Most of the energy is then directly transferred from the forcing scale to the jet scale through direct interaction with the average flow. This process, which transfers energy non-locally in Fourier space, is not a spectral
cascade anymore \citep{huang1998two,Bouchet_Nardini_Tangarife_2013_Kinetic_JStatPhys,bakasioannou2013}.

\begin{figure}[htpb]
\centerline{ \includegraphics[width=12cm]{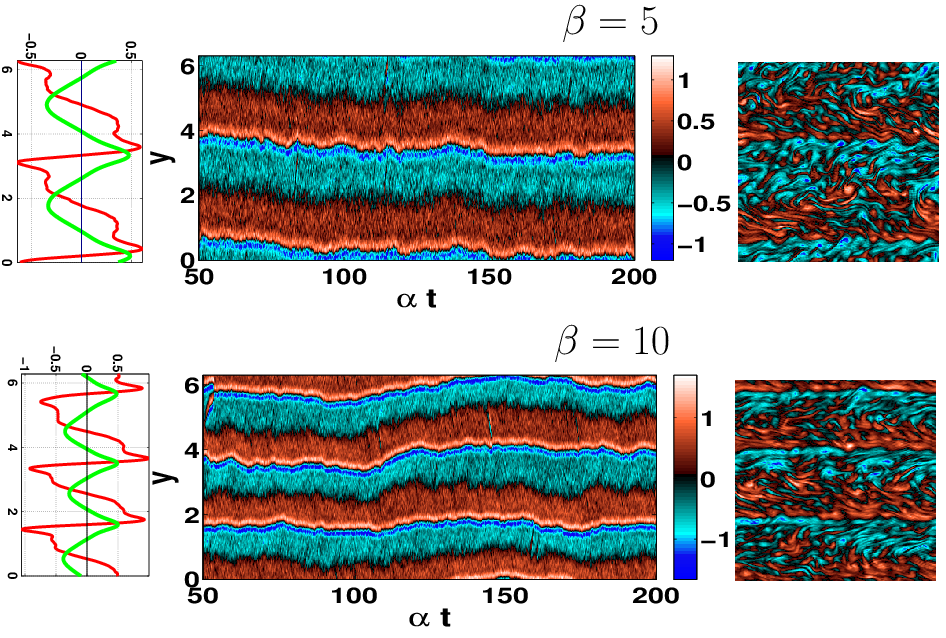}}
\caption{Jets sampled in direct numerical simulations. Left panels show local time averages of the zonally averaged vorticity (red curve) and velocity (green curve) as a function of latitude $y$ ($y$ is the vertical axis), for $\alpha = 1.2 \cdot 10^{-3}$.
Middle panels show the corresponding Hovm\"oller diagrams of the vorticity $\omega$ (the color represent the value of the zonally averaged vorticity as a function of time and latitude).
Right panels show one of the corresponding instantaneous snapshots of the vorticity field.
The first line shows these figures at $\beta=5$, the second line shows these figures at $\beta=10$.}
\label{dns}
\end{figure}

\begin{figure}[htpb]
\centerline{ \includegraphics[width=8cm]{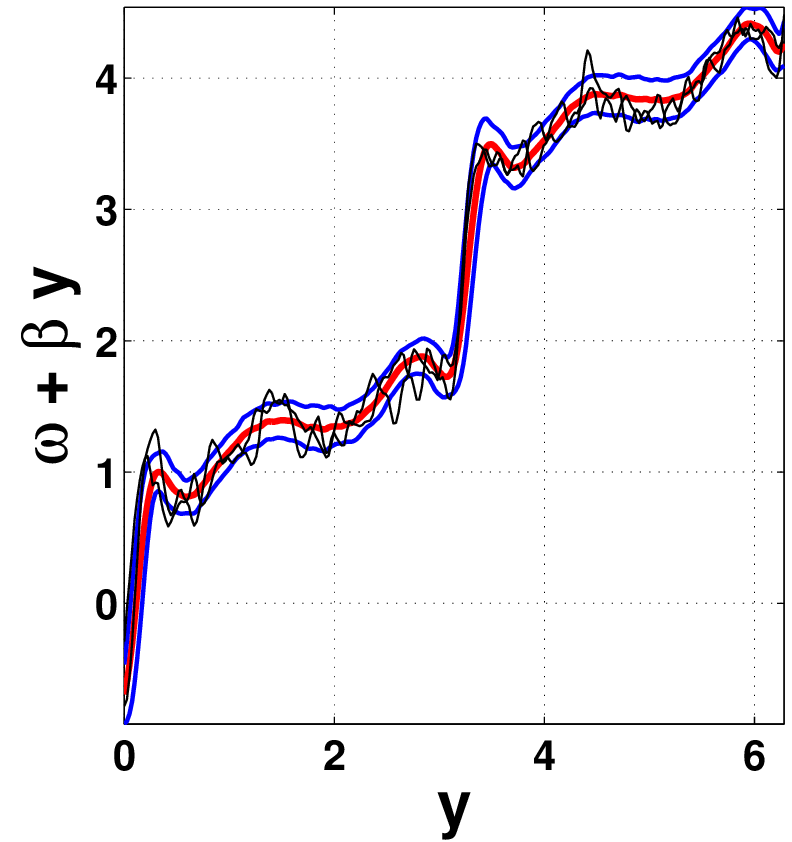}}
\caption{Behavior of the zonally-averaged potential vorticity $\omega + \beta y$ with the same parameters than in Fig.\ref{dns} ($\beta =  5$).
The red curve corresponds to the time mean value for $\alpha t \in [100,150]$, the blue curves to the corresponding $\pm$ standard deviations, and
the black curves to two different times chosen arbitrarily.}
\label{jaspv}
\end{figure}

In the following, we will be interested in this strong jet regime, which is relevant for Jupiter. Except if otherwise stated, all the computations of this paper will be performed with the parameters $\alpha = 1.20\ 10^{-3}$, $\nu = 1.5\ 10^{-8}$, and using a stochastic force with a uniform spectrum in the wave number band $|k| \in [14,15]$. The related forcing scale will be always well below the Rhines scale. Figure \ref{dns} illustrates that although the flow is clearly turbulent, in the statistically stationary state the zonal jets are very stable on timescale of the order of $10^5$ turnover time or more. If $\alpha$ would be decreased, these jets would become even more stable. The averaged vorticity has a saw-tooth profile: it is composed of homogenized potential vorticity areas where the potential vorticity $\omega + \beta y$ is nearly constant (and for which the vorticity decreases approximately like  $-\beta y$ across the basin), separated by small areas of abrupt increase of the vorticity (or potential vorticity).
This PV staircase phenomenology has been first described by \cite{marcus1998model}
and is only approximately valid \citep{Danilov}. We illustrate in Fig.~\ref{jaspv} how the typical potential vorticity mean profile behaves, using the same solution
than in Fig.~\ref{dns}.
For the velocity profile, homogenized potential vorticity areas correspond to westward jets, with a local quadratic velocity profile with a curvature approximately equal to $\beta$, while the jumps in vorticity give rise to cusps for the velocity close to eastward jets. This phenomenology has been observed in many simulations for barotropic flows.  The top panels of Fig.~\ref{dns} show a state with two alternating jets. There are actually four jets: two eastward jets with local maxima of the zonally averaged velocity and two westward jets with local minima of the zonally averaged velocity. We stress also that, on the Hovm\"oller diagram, the jet extrema are located on the black area (close to zero vorticity lines). The eastward jets are the black areas with negative (blue) vorticity on the south and positive (red) vorticity on the north.
Related to the PV staircase phenomenology, the black areas for eastward jets are very thin line, while the black areas for westward jets have a broader extent.
In the doubly-periodic domain of this study, the number of eastward jets has to be equal to the number of westward jets, and jets will come by pairs. We will call a state with $K$ alternating jets, a $K$-jet state, i.e. a state with $K$ eastward and $K$ westward jets.

\section{Rare transitions between states with different jet numbers}\label{sams}

\subsection{Bistability, rare-transitions and hysteresis between states with two and three alternating jets}\label{sbim}

As stressed in \cite{dritschel2008multiple}, assuming the PV staircase phenomenology with a westward jet curvature of order $\beta$ is sufficient to roughly determine the number of jets: its order of magnitude is then given by the ratio of the domain size divided by the Rhine scale, or equivalently in our non-dimensional units $\sqrt{\beta}$. Figure \ref{dns} indeed shows that increasing $\beta$ increases the number of jets.

Compatible with this phenomenology, one expects to see transitions from $K$ alternating jets to a state with $K+1$ ones when $\beta$ is increased. As there is no symmetry breaking in this process, one may expect these transitions to be first-order ones with discontinuous jumps of some order parameters. In situations with discontinuous transitions when an external parameter $\beta$ is changed, one expects for each bifurcation multistability ranges $(\beta_1,\beta_2)$ in which two (or more) possible states are observed depending on the initial conditions, and for which extremely rare transition from one state to another may occur due to either external or internal fluctuations. Such a bistability phenomenology has first been observed in \cite{Bouchet_Simonnet_2008}. The aim of this paper is to go much further in the study of the structure of the attractors and the description of the spontaneous transitions between them. \\

\begin{figure}[htpb]
\centerline{ \includegraphics[width=12cm]{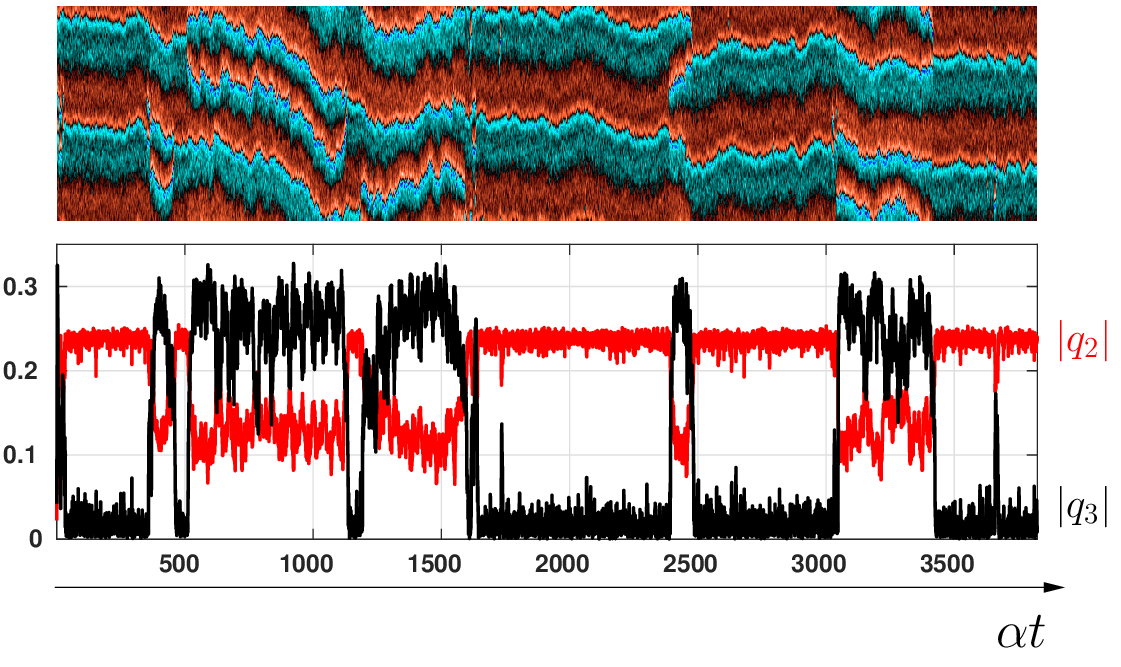}}
\caption{Lower panel: timeseries the moduli of the  first zonal Fourier components ($|q_n|$ with $q_n = \frac{1}{|\Omega|}\int_\Omega \omega(x,y)~{\rm e}^{i n y} \ {\rm d}x{\rm d}y$) for $n=2$ (red) and $n=3$ (black). The corresponding Hovm\"oller diagram for the zonally averaged vorticity is shown in the upper panel ($\alpha = 1.2 \cdot 10^{-3}$ and $\beta = 5.2$).}
\label{bimodns}
\end{figure}

Figure \ref{bimodns} shows that the bistability occurs for fixed values of the control parameters and a very wide range of initial conditions. We compute the Fourier components of the vorticity, with zonal wave number $(0,n)$, defined as $q_n = \frac{1}{|\Omega|}\int_\Omega \omega(x,y)~{\rm e}^{i n y} dx~dy$, for $n=2$ and $n=3$. Their moduli  $|q_n|$ are relevant order-parameter, for instance $|q_2|$ will be large for a two-alternating jet state and small otherwise. In Fig.~\ref{bimodns}, one observes five spontaneous $2 \to 3$ transitions and five $3 \to 2$ transitions for a duration of about $10^6$ turnover times. The fluctuations of the three alternating jet states are noticeably larger than the ones for the two alternating jet states. This corresponds to some internal dynamics between several three alternating jet states, as will be explained in section \ref{internald}.

\begin{figure}
\centerline{\includegraphics[width=10cm]{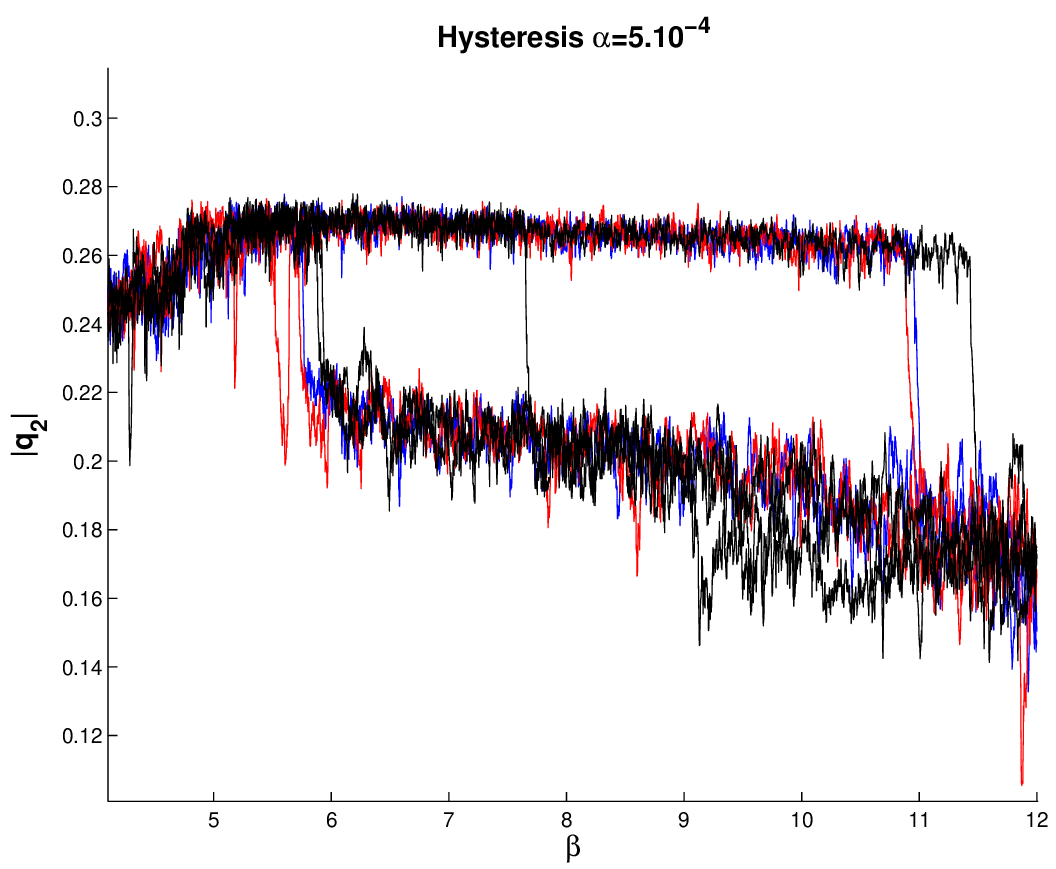}}
\caption{We performed four independent hysteresis experiments, by increasing and then decreasing slowly the value of $\beta$. For the four experiments, each of the four curves represents $|q_2|$, the modulus of zonal Fourier component with wavenumber two, as a function of $\beta$ ($\alpha = 5 \cdot 10^{-4}$ and
$\dot \beta = \alpha/100$).}
\label{HY}
\end{figure}

We finally perform an hysteresis experiment by varying adiabatically (very slowly) the parameter $\beta$  at rate $\dot{\beta}=\frac{\alpha}{100}$, in the range $\beta \in [4,12]$, with a smaller value of $\alpha=5\cdot 10^{-4}$. Fig.~\ref{HY} shows different numerical experiments with independent realizations of the noise. This figure displays the modulus of $|q_2|$ as a function of $\beta$ in three of these experiments: the upper branch of $|q_2|$ corresponds to the two alternating jet state and the lower branch to the three alternating jet state. It shows bistability in the range $\beta \in [6,11]$, with transitions occurring after a transition time $T\lesssim 10^{5}$: with this slowly varying $\beta$, transition events are not uncommon. However, performing direct numerical simulations for this value of $\alpha$ will not show transitions. The reason is simply that, as explained in the next sections, the probability of seeing such transitions becomes then much too small. Based on the hysteresis curve, we choose $\beta = 8$ as a reference value for future studies, unless otherwise specified.

\subsection{Nucleation and coalescence of jets}\label{samsb}
\begin{figure}
\centerline{\includegraphics[width=12cm]{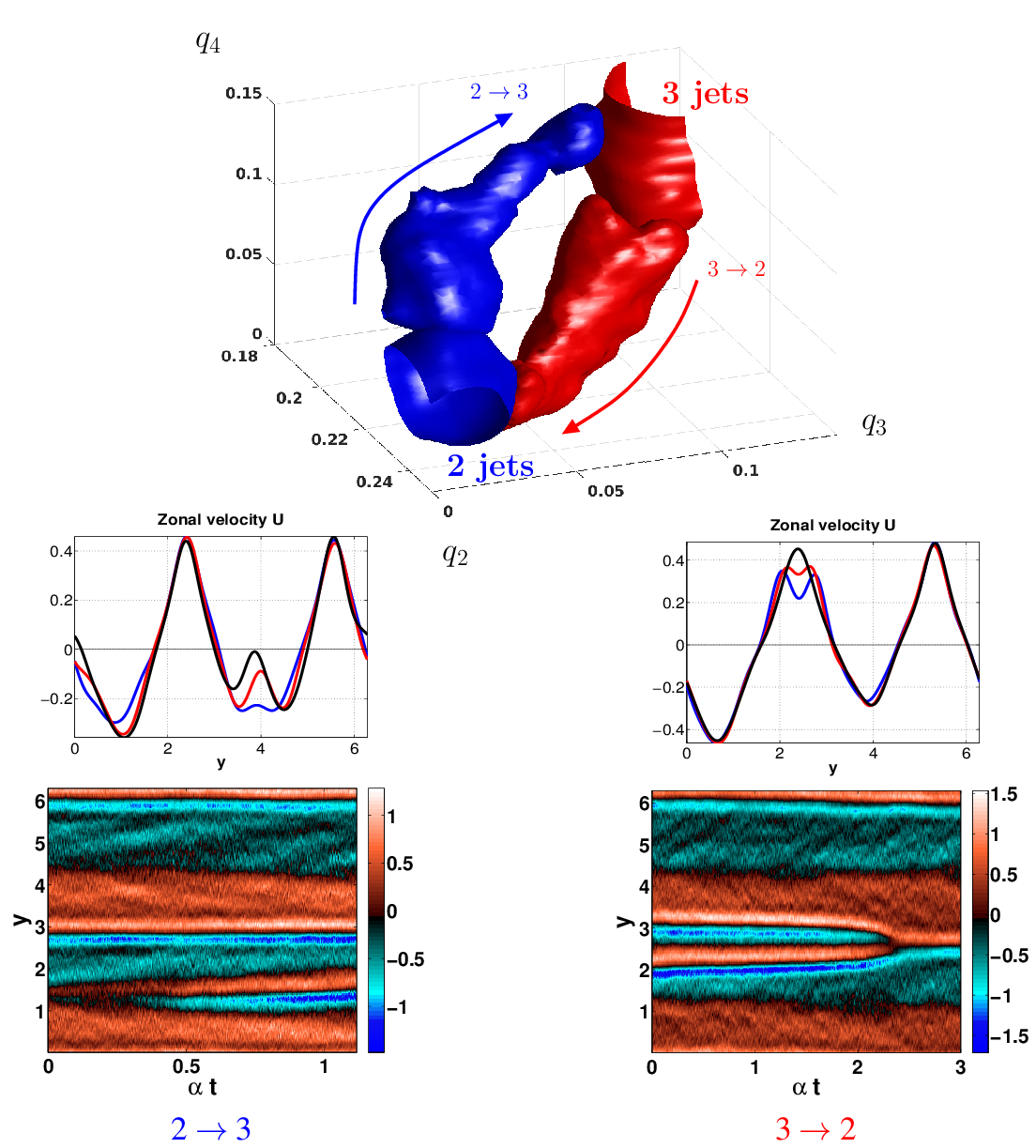}}
\caption{The upper panel shows the level set of the reactive trajectories in the space $|q_2|,|q_3|,|q_4|$. The blue and red reactive tubes
contain 80\% of the $2 \to 3$ transitions and $3 \to  2$ transitions, respectively. Each of these tubes gather about $1000$ reactive trajectories obtained
using the AMS algorithm, for $\alpha = 10^{-3}$ and $\beta = 8$. Middle left and right panels show the zonally averaged zonal velocity as a function of latitude, 3 curves at 3 different times (blue first, then red, then black), during the $2\to 3$ and $3 \to 2$  transitions, respectively. On the left, we observe the nucleation of a new jet at the westward tip of an existing jet.  On the right, we observe the merging of two eastward jets. Lower left and right panels are the two corresponding Hovm\"oller diagrams for the zonally averaged vorticity.}\label{TR}
\end{figure}

We discuss now the spontaneous $2 \to 3$ and $3 \to 2$ transitions, as seen in Fig.\ref{bimodns}. While we could not obtain many transitions with direct numerical simulations, using the rare event AMS algorithm, we were able to collect several thousands of
 $2 \to 3 \to 2$ transitions. The dynamics of those transitions are represented in Fig.\ref{TR}, in the space spanned by the order parameters $|q_2|,|q_3|$ and $|q_4|$. The remarkable
result is that these trajectories seem to concentrate in the phase space. The blue tube in Fig.\ref{bimodns} contains 80\% of the $2\to 3$ transitions whereas the red tube contains 80\% of the $3 \to 2$ transitions. The concentration of trajectories that lead to a rare events close a predictable path, is called an instanton phenomenology (see for instance \citep{Laurie,Grafke,BLZ,dematteis2019experimental} for turbulent flows). Figure \ref{TR} illustrates this remarkable instanton phenomenology for the  $2 \to 3 \to 2$ transitions. For this parameter regime, $2 \to 3$ and $3 \to 2$ transitions  occur with a return time approximatively equal to $10^{8}$. Note that it is a very rough estimate based on a single AMS realization. We did not search for more precise estimates here, as we are interested mostly on transitions dynamics. In principle, one should use more realizations with different reaction coordinates (see Appendix B).

The middle-left panel of Fig.\ref{TR} shows that the nucleation of a new pair of alternating jets proceeds through an inversion of the velocity curvature of one of the parabolic westward jet in the exact middle between the two adjacent eastward jets. It takes the form of a small bump in the zonal velocity profile.
This structure indeed nucleates from two bands of positive and negative vorticity inside a narrow band of zero vorticity around the westward jet. Such a nucleation is highly improbable, as the new vortex bands are initially too narrow to be stable. Then just by chance, one can have this band surviving long enough to grow until it reaches a stable state.  However, once a critical size is reached, the band becomes stable
and grows in size. The three jets slowly equilibrate and separate apart. The new jet can relax either to the north or to the south of the initial parabolic jet where it has nucleated. We will describe this in more details in the next section.
In the following, we will call a transition $2 \to 3$, a {\it nucleation}.

The lower-right panel of Fig.\ref{TR} shows how two jets can merge, which can be interpreted as the disappearance of one pair of alternating jets. The phenomenology is again rather simple. Interestingly, if one reverses both the time and the sign of the velocity, one would indeed obtain a nucleation
from a cuspy (westward) jet. Moreover, the resulting two alternating jet state just after the two eastward jets have merged, is slightly asymmetric (see zonally averaged zonal velocity for the $3 \to 2$ transition in Fig.\ref{TR} and section \ref{internald} hereafter). This state then relaxes to a perfectly symmetric two-jet configuration over a timescale of order $1/\alpha$ (not shown). We call the jet merging $3 \to 2$ transition a {\it coalescence}.

\subsection{Barotropic adjustment and Rayleigh-Kuo criterion for zonal jets and during transitions}\label{samsc}

\begin{figure}[hptb]
\centerline{\includegraphics[width=7cm]{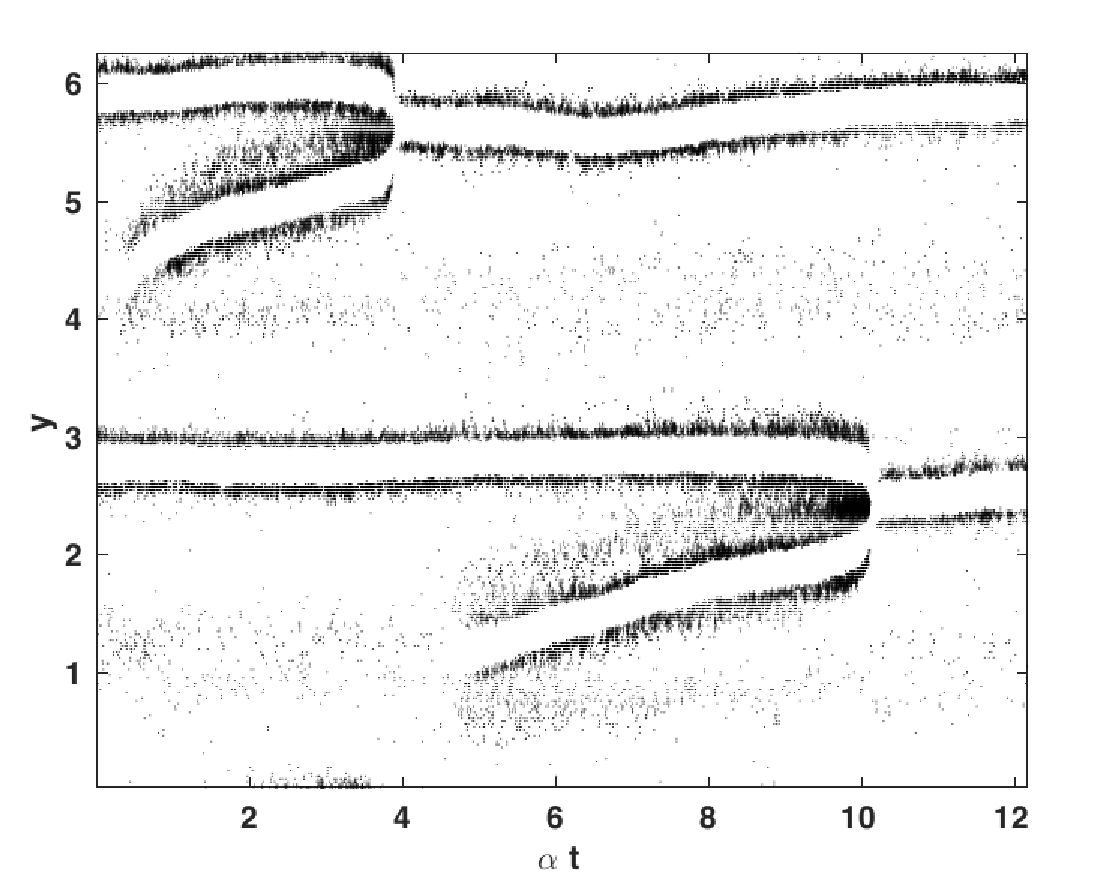}}
\caption{\label{fig:Rayleigh_Kuo}
The Rayleigh--Kuo criterion for an example with two nucleations and two mergers. This is a Hovm\"oller diagram where black dots are plotted each time $RK=U''-\beta$ is positive ($\alpha=2.2\ 10^{-4}$ and $\beta=5.5$).
}
\end{figure}
Are rare transitions between attractors driven by hydrodynamic instabilities? This is a very natural fluid mechanics question that we address now. It is clear from all the Hovm\"oller diagrams, on Figs~\ref{dns} and \ref{TR} and on movies of the dynamics (not shown), that the time scale for the transition dynamics is of order $1/\alpha$, both during a nucleation of new jets or during a coalescence. This is also true for all other transitions we have observed. This phenomenology  thus excludes a fast instability in the barotropic equation (\ref{eq:barotropic}), developing on time scales of order one. As a consequence, if some instability occurs in equation (\ref{eq:barotropic}), the flow has to be marginally unstable with an instability rate scaling at most like $\alpha$ when $\alpha \downarrow 0$.

Such an observation of marginal instability is related to the classical hypothesis of barotropic
adjustment of eddy-driven jets (see \cite{read2020turbulent,read2020baroclinic} and references therein). An eddy-driven jet with a time-scale
separation between eddy dynamics and zonal jet time scale, should
be unstable in order to transfer meridionally zonally-averaged momentum,
while it should be at the same time stable if quasi-stationary zonal jet
states are observed for very long times. This apparent contradiction
leads to the hypothesis of barotropic adjustment: the state of the
system should be marginally stable (or unstable) in order to fulfill
these two seemingly contradictory requirements. The relevance of this
adjustment hypothesis has been recently discussed in the context of
Jovian planets and for a hierarchy of models \citep{read2020turbulent,read2020baroclinic}.
The interesting remark we want to stress here is that this barotropic adjustment mechanism actually takes place for the statistically quasi-stationary jets, as might be expected, but perhaps more unexpectedly, also takes place during the periods of relaxation towards these quasi-stationary states and during the periods of "instability" that lead to coalescences and nucleations of jets.

In order to have a more precise and quantitative insight on this issue, we look at the Rayleigh--Kuo criterion $RK=U''-\beta$,   where $U''$ is the second derivative with respect to $y$ of the zonally averaged zonal velocity $U$. It is well known that, for a zonal flow $U(y)$, a necessary condition for an hydrodynamic instability to occur is that the Rayleigh--Kuo criterion changes sign \citep{KUO}. Figure~\ref{fig:Rayleigh_Kuo} is a Hovm\"oller diagram of the sign of the Rayleigh--Kuo criterion: we draw a black point each time the Rayleigh criterion is negative. This picture is drawn from an example of a $2\to3$ transition  obtained with the adaptive multilevel splitting algorithm ($0 < \alpha t < 1.3$) followed by a free dynamics (Eq. (\ref{eq:barotropic}) without selection) ($1.3 < \alpha t < 12$), with the parameters  $\alpha=2.2\ 10^{-4}$ and $\beta=5.5$. The eastward jets are localized inside the thin white bands delimited by the area with dense black dots. The westward jets are localized in the area with lots of scattered black dots, but with a much lower density than at the flank of eastward jets. The picture then shows subsequently: a nucleation of new jets at the location of a westward jet ($0 < \alpha t < 1.3$), the northward drift of one of these new jets until it reaches another eastward jet and coalesces ($1.3 < \alpha t < 4$), a transient  state with two alternating jets ($4 < \alpha t < 4.9$), the nucleation of new jet and the drift northward of one of them, until a new coalescence ($4.9 < \alpha t < 10.2$), and a stable state with two alternating jets ($10.5 < \alpha t < 12$).

We first describe in detail Fig.\ref{fig:Rayleigh_Kuo} during the period with a stable state with two alternating jets ($10.5 < \alpha t < 12$). We note that most of the time, the Rayleigh-Kuo criterion is positive, meaning that the overall structure is stable. However, from time to time, the Rayleigh--Kuo criterion changes sign on some specific points of the domain. During that period, the velocity profile $U$ looks very much like the one of the two-jet state. On this curve, we see a characteristic shape for eastward jet velocity profile, with a characteristic cusp shape. At the edge of the eastward jet, the cusp is associated with a very strong negative curvature and thus a
very low value of $RK=U''-\beta$, explaining the white band without black dots on Fig.\ref{fig:Rayleigh_Kuo}. On each flank of the jet, the velocity profile has a concavity change, where $RK$ changes sign very often, explaining the two dense black dot bands surrounding the eastward jet. On all the other parts of the flow, $RK$ remains much of the time positive, except close to the westward jet, where $U''$ is very close to $\beta$ and $RK$ changes sign rather often. This explains the less dense band of black dots close to the westward jet edge. We stress also that the shape of eastward jet cusps seems to depend significantly on the dissipative mechanism. Indeed, Figs \ref{dns} and \ref{TR} show that, for larger values of $\alpha$ for which hyperviscosity effects are negligible, the cusp is rather rounded, while other simulations with smaller values of $\alpha$ (not shown), for which hyperviscosity effects become important, the velocity profile has a strong concavity change on the flank of the jet. In a recent theoretical work \citep{Woillez}, it has been established that: i) eastward jets should form a cusp, with a width of order $1/K$ where $K$ is the typical random force wavenumber, and which specific profile depends on the dissipative mechanism, ii) the mechanism preventing the growth of westward jet is a marginal instability of the velocity profile appearing when the positive curvature of the jet is of order $\beta$. Our numerical observations are thus in line with this theoretical work.

Using Fig.\ref{fig:Rayleigh_Kuo}, we now describe the jet nucleations and coalescences, from the point of view of the Rayleigh--Kuo criterion. First we observe that during the nucleation processes, we do not see any change of sign for the Rayleigh--Kuo criterion that would qualitatively differs from the stable jet situation. Second, during the coalescence, a black band appears at the level of the westward jet squeezed between two eastward jets, just before the coalescence. In this area, although we cannot conclude from this figure, the Rayleigh--Kuo criterion may change sign. As the Rayleigh--Kuo criterion is only a necessary condition for instability, we can not conclude from this figure if a very localized and very low rate instability may occur during the coalescence. However, if such an instability actually occurs, its rate has to be very small, as the typical time scale for the change of the structure is very long, of order $1/\alpha$.

We conclude that the flow is constantly marginally stable close to westward jets, in accordance with a barotropic adjustment picture. This is the case for stationary states, as we already analyzed and discussed theoretically in~\cite{Woillez}, but this is also the case during both the coalescence and the nucleation transitions. This marginal stability is thus not specifically related to transition trajectories, as it also occurs for stationary states. We conclude that this marginal stability (or instability) is not by itself the driver of the transitions.

In this section, we have addressed the hydrodynamical stability of the flow. We have seen that no hydrodynamical unstability occurs, although marginal stability is always present. A different question is wether the slow effective dynamics that describes the flow on longer time scale should be unstable or not. It should be unstable, at some point, because of the observed saddle point phenomenology. When transitions occur, in order to pass the saddle point, the noise has to compensate the  instability of the slow effective dynamics. On the slow timescale, the relevant noise is the one produced by the fluctuations of the divergence of the time average of the Reynolds stress. We discuss this in the next section.

\subsection{Saddle points}\label{saddle}\label{samsd}
In the following, we will use the language of dynamical system theory plus weak noise, in order to describe the phenomenology of the attracting states and the transition dynamics between them. This dynamical system discussion applies to the slow effective dynamics of the zonally averaged zonal
velocity $U(y,t)=\frac{1}{2\pi}\int_0^{2\pi} u(x,y,t)~dx$. Indeed, in the regime of small values of $\alpha$, as argued in \cite{Bouchet_Nardini_Tangarife_2013_Kinetic_JStatPhys}, we expect the slow relaxation dynamics of zonal jets to be described by a deterministic effective equation that can be obtained through a second cumulant closure.  This kinetic theory has been extended to take into account marginally stable modes (\cite{Woillez}) as it occurs at the tip of westward jets (see previous sections).

Besides such deterministic effective equations, the transitions from one attractor to another are expected to be triggered by a noise, due to the fluctuations of the Reynolds stress divergence, which properties can be computed using large deviation theory as explained in \cite{bouchet2018fluctuations}. Based on these theoretical works, we expect the effective dynamics to follow.
\begin{equation}\label{effeq}
\partial_\tau U = F(U) + \sqrt{\alpha} \sigma(U,\tau),
\end{equation}
with the rescaled time $\tau = \alpha t$, the average of the divergence of the Reynolds stress divergence $F(U)$, and a white noise $\sigma(U,\tau)$ related to the fluctuations of time average of the divergence of the Reynolds stress.

Interesting formulas for $F$, implying local derivatives, were proposed by \cite{srinivasan2014reynolds}, or by \cite{laurie2014universal} in a different but related context. As explained in \cite{Woillez}, such local formulas and others can be justified in the limit when the forcing scale is much smaller than the jet width, and lead to interesting dynamical consequences. However this is only valid in flow areas with strong enough shear. Such local formulas would thus not be valid globally for the flow, for instance close to jet tips, even if the spatial scale separation would be present. We conclude that, while the theoretical works clearly justify the existence of equation (\ref{effeq}) under some natural hypothesis, the formulas for $F$ and $\sigma$ are not explicit in general. They might be computed numerically. In the absence of analytical expression and easy numerical computation from the theory, the aim of this work is to directly verify the dynamical and qualitative consequences of  (\ref{effeq}) without computing explicitly $F$ and $\sigma$.

The attractors and saddle points should be thought as attractors and saddle points of the deterministic dynamics $\partial_\tau U = F(U)$.
Although there is no explicit knowledge of $F$, such attractors and saddle points can neverveless be directly observed in fully turbulent direct numerical simulations, or in fully turbulent simulations using the AMS algorithm. In a fully turbulent simulation, a saddle point belongs to a transition trajectory. Moreover, when adding small perturbations it can either relax to one attractor or to the other depending on the perturbation. Those two properties are enough to characterize them numerically, as we better explain in appendix B.c.

For the value $\beta = 8$, the hysteresis curve in Fig.\ref{HY} suggests that the system is multistable with at least two attractors. For dynamical systems
with weak noise, in the simplest cases, the transitions from one attractor to another must go through a saddle point which belongs to the common
boundary of the two basins of attraction. In more complex situations, the transitions could rather proceed through more complex structures, for
instance saddle limit cycles or saddle strange attractors. These scenarii where two basins collide are called {\it crisis bifurcations} \citep{crisis}.
As we describe in the following, we have detected saddles points only for the dynamics of the $2 \to 3$ and $3 \to 2$ transitions.

Using AMS results, we are able to easily find these saddle states, using the methodology described in appendix B.c. They are shown in Fig.\ref{nuksad} for the nucleation. As mentioned before, the new
jet is created from the parabolic jet at the middle of the two westward jet. This new jet can drift either to the north or to the south. This is illustrated by the
presence of two different ensembles in the AMS reactive trajectories giving two different saddles. The difference is small however and it is possible
that in other parameter regimes, these two saddles collapse into a single one at the exact middle of the two westward jets. This is discussed in the next sections.
Note that, in addition, one must add two other saddles related to the nucleation at the other westward jets. Moreover,
it is possible that one observes direct transitions $2 \to 4$ involving four possible double nucleations
(north north, north south, south north and south south). We have indeed observed such transitions (not shown), but we
have decided not to discuss higher-order transitions involving several transitions at the same time, that is transitions of codimension larger than one.

\begin{figure}
\centerline{\includegraphics[width=14cm]{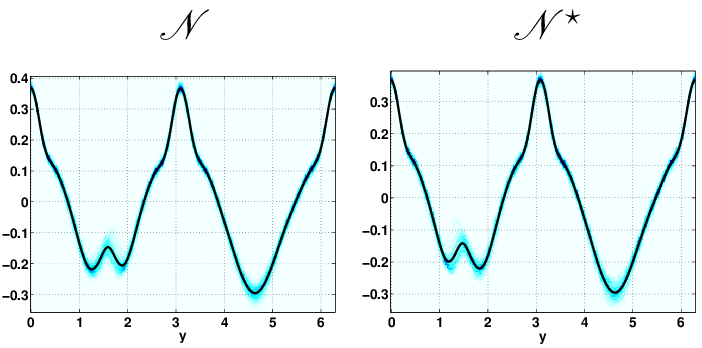}}
\caption{Zonally averaged velocity as a function of latitude, $U(y)$, for the nucleation saddles. Each black curve is obtained by the ensemble average of the output of about 800 AMS runs ($\alpha = 10^{-3}$ and $\beta=  8$). Please note the slight asymmetry of each of the saddles, close to the westward tip of the nucleating jet. For each value of $y$, we have superimposed the histograms of the $U$ values of the 800 AMS states (light blue), in order to feature the spread of the ensemble values.}\label{nuksad}
\end{figure}

\begin{figure}
\centerline{\includegraphics[width=7cm]{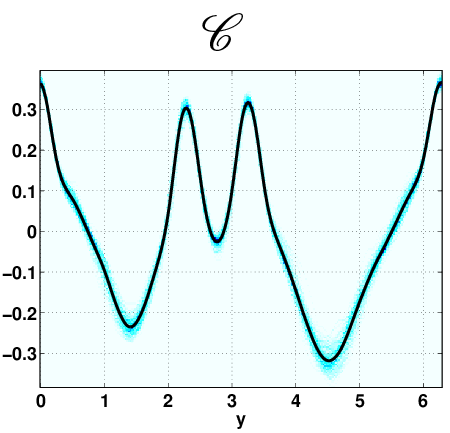}}
\caption{Zonally averaged velocity as a function of latitude, $U(y)$, for the coalescence saddle. Each black curve is obtained by the ensemble average of the output of about 200 AMS runs
($\alpha = 10^{-3}$ and $\beta=  8$). For each value of $y$, we have superimposed the histograms of the $U$ values of the 800 AMS states (light blue), in order to feature the spread of the ensemble values.}\label{coasad}
\end{figure}

A similar study of the saddle points for the transitions $3 \to 2$ is discussed in Fig.\ref{coasad}.
If the distance between the two closest jets is below a certain critical distance (approximately one sixth of the domain width), they start to be attracted to each other
until they merge into a single jet.
On the contrary, if their distance is above this critical value, the whole system relaxes to a stationary three-jet state.
This critical distance appears to be rather sensitive to the value of $\beta$.
This distance increases as $\beta$ decreases and decreases for larger values of $\beta$. This is consistent with the idea that for a regime
where only the two-jet states dominate, say for smaller values of $\beta$, the three-jet states will be easily destabilized though the coalescence process.
One may ask, why only one saddle is found ? The reason is simply related to the initial conditions used in the AMS. In fact,
there are as many coalescence saddles as nucleation saddles. This point is clarified in section \ref{internald}.\\

Two natural scenario can be considered for the transitions between states with various number of jets either: i) an hydrodynamic instability of the base flow $U$ or ii) instabilities in the effective dynamics that describe the evolution of the flow on longer time scales after averaging. \cite{constantinou2014emergence} shows clear evidence of scenario ii) for some range of parameters. From the discussion in this section and the previous one, we see that the dynamics described in this paper follow a third scenario, with both marginal stability for the hydrodynamic equations and instability of the effective dynamics. Indeed, as explained in the previous section our flow is constantly marginally stable. This marginal stability might or might not be related to the transition dynamics\footnote{Although we have not studied carefully this point, we believe that it is likely that marginal modes at westward jets are involved in the nucleation of new jets, but not involved for the coalescence of two jets}. Then one needs to generalize effective theories in order to describe long term dynamics of the zonal jet in order to take into account marginally stable states, for instance following \cite{Woillez}. Then,
the existence of saddle points requires instabilities of the effective dynamics. Our third scenario has thus flavors of both scenario i) and ii).

\subsection{Instantons}
Figure \ref{TR}  strongly suggests the existence of an instanton-driven dynamics. By this, we mean that jet nucleation or coalescence events are concentrated around a predictable path, called instanton. These paths are composed of a rare noise-driven fluctuation path from the two alternating jet attractors toward a saddle point (for instance), then of a typical relaxation path from the saddle point to the three alternating jet attractor. If our hypothesis is correct, as $\alpha \to 0$ the
reactive tubes shown in Fig.~\ref{TR} should collapse to a single instanton path (see Appendix A).
By an abuse of language, we will assume that this is the case and we identify the instanton with the reactive tube.
This is justified and discussed in section \ref{ldp}, when we vary $\alpha$.

\begin{figure}
\centerline{\includegraphics[width=15cm]{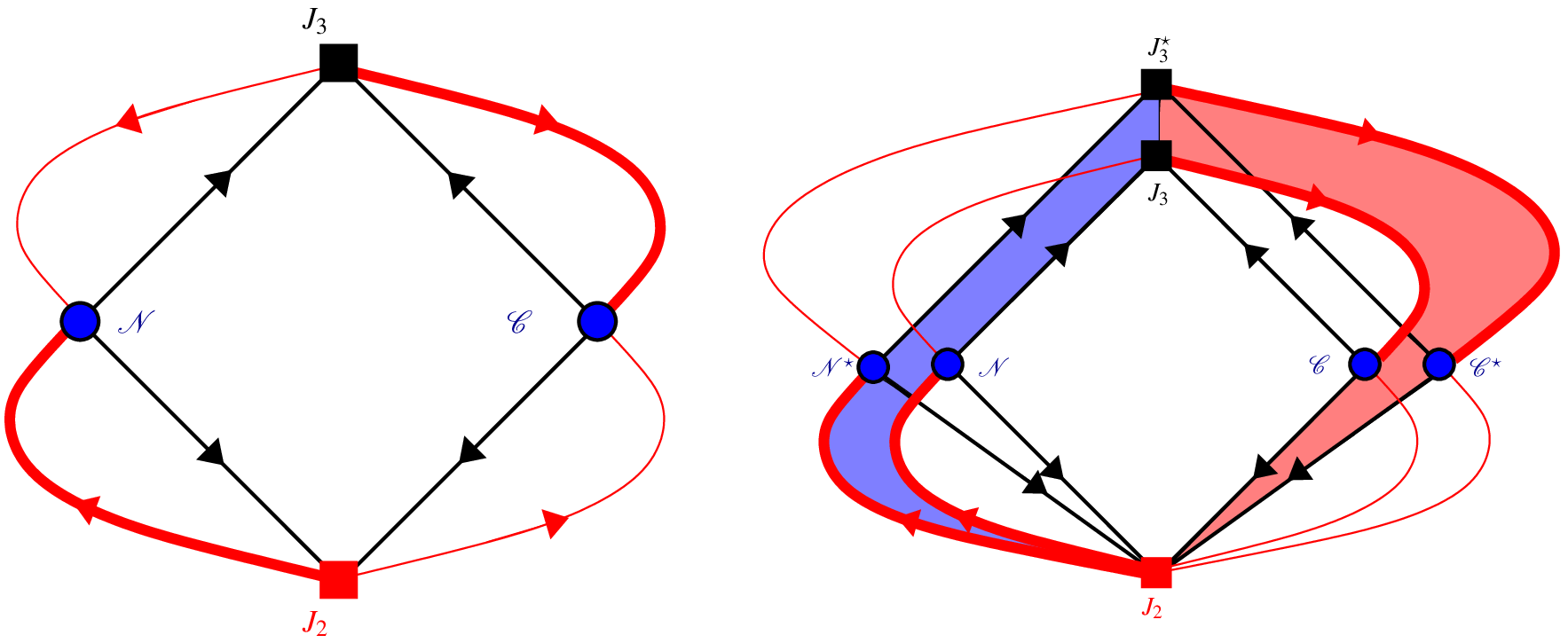}}
\caption{Schematic representation of the instantons involved in the transitions $2 \to 3 \to 2$. Left figure is a simplification of
the right figure, ie without representing the dual mirror pair of states.
$J_2$ is the stationary two-jet state, $J_3$ and $J_3^\star$ are the two mirror-symmetric three-jet states obtained from the two nucleation saddles called
${\cal N}$ and ${\cal N}^\star$ (see Fig.\ref{nuksad}). The coalescence saddles are called ${\cal C}$ and ${\cal C}^\star$ (see Fig.\ref{coasad}). The black arrows represent
the relaxation trajectories from the saddle to the attractors (with zero action; see Appendix A). The red curves correspond to the reactive part of the instantons (with nonzero action). The colored area
inside the bold curves is what is shown in instanton tubes of Fig.\ref{TR}, that is half of the actual instanton structure. }\label{insts}
\end{figure}

The important point is that two different saddles are involved in the $2\to 3$ and $3 \to 2$ transitions, and not just one. The consequence
is a more subtle structure which is not seen in the two crude Fig.\ref{TR}. This is explained by the schematics of Fig.\ref{insts}. Let us denote the two alternating jet attractor and the three alternating jet attractor $J_2$ and $J_3$ respectively.
In Fig.\ref{insts}, the nucleation instanton is the red curve starting from $J_2$ and going to ${\cal N}$
together with its relaxation part (black curve) from ${\cal N}$ to $J_3$.
Similarly, the coalescence instanton is the red curve starting from $J_3$ to ${\cal C}$ and its relaxation part
(black curve) goes from ${\cal C}$ to $J_2$. The relaxation curves (black curves) correspond to trajectories having zero action in the asymptotic
limit since they correspond to the natural dynamics of the system.
Figure \ref{nuksad} shows that the left schematic of Fig.\ref{insts} has moreover a mirror equivalent shown to the right.
The mirror states are denoted here with a star for convenience.

Moreover, to each instanton is associated an anti-instanton going through the same saddle.
The pair instanton anti-instanton forms a so-called figure-eight pattern in phase space.

The time asymmetry does not only translate into a geometrical asymmetry of the pair instanton anti-instanton, but the transition probability of
corresponding reactive trajectories is different as well.  It appears that in this model, the asymmetry is very strong especially in terms of transition
probabilities: the ``anti-transition" probabilities are several order of magnitude smaller than the transition probabilities.
For instance, the "anti-coalescence" instanton corresponds to a path going from $J_2$ to ${\cal C}$ (thin red curve) and then from
${\cal C}$ to $J_3$ (relaxation black curve).
It requires an inversion of the curvature on the top of one of the eastern cuspy jets followed by a separation of the two small jets which are created off the cusp.
Such events are of course very rare but can be detected by AMS. The probability of an inversion of the cusp curvature is estimated by AMS to be $O(10^{-9})$.
However, the nucleation of an eastward jet leading to the configuration shown in Fig.~\ref{coasad} has a considerably smaller probability, estimated by AMS to be less than $O(10^{-30})$.

We describe in the next section the internal effective dynamics of the state with three alternating jets.
\section{States with three alternating jets and their internal dynamics}\label{internald}
The previous results indicate the existence of at least two types of states having three alternating jets, due to different possible nucleations (see Fig.\ref{nuksad}) or coalescences (see Fig.\ref{coasad}).
It was previously referred to as a mirror symmetry without being too precise on what symmetry was involved.
We show next that the reality is more complicated, with the description of several different states with three alternating jets, and a slow dynamics between them. This explains in particular the large amplitude of the fluctuations of the three alternating jet states, as seen in Fig.\ref{bimodns}.
\subsection{New representation of the jet attractors and their dynamics}
In order to describe this structure, we need a more precise description of the jet profiles (the zonally averaged zonal velocity). If one measures precisely the distance between the jet tips, one observes some slight asymmetries in the three alternating jet states. Indeed, the attractors do not equilibrate
to an equidistant jet solution, see for instance the three alternating jet states in Fig.\ref{bimodns}. This asymmetry is robust when $\beta$ is changed. Based on this remark, it is relevant to characterize
these states by the distance between the eastward jet tips for instance. These distances give three numbers we call $\sigma_1,\sigma_2,\sigma_3$. Our convention is to
divide these quantities by the dimensionless domain width $2 \pi$ so that $0 \leq \sigma_i \leq 1$. Due to the periodicity in $y$, one has the constraint
\begin{displaymath}
\sigma_1 + \sigma_2 + \sigma_3 = 1.
\end{displaymath}
Given this constraint, in the following, we prefer to consider the most economical representation and use only two of these quantities, say $\sigma_1$ and $\sigma_2$. Looking at
the Hovm\"oller diagram of Figure \ref{bimodns} for instance, one observes some slow drift of the solution in the meridional direction.
We therefore choose one of the eastern jet as a reference moving frame, and compute $\sigma_1$ and $\sigma_2$ from this reference jet. $\sigma_1$ is the distance between this reference jet and the next eastward-jet tip, northward. $\sigma_2$ is the distance between this new jet and the next eastward-jet tip, northward again. This procedure is for measurements only and requires to always track the same reference jet at any time. A state with three alternating jets will be thus described by $(\sigma_1,\sigma_2,\sigma_3=1-\sigma_1-\sigma_2)$.

The values taken by $\sigma_1$, $\sigma_2$ and $\sigma_3$ are always very close to three fixed values $\sigma_1^*$, $\sigma_2^*$ and $\sigma_3^*$ which are all different from $1/3$ and different from each others. On figures \ref{den}, \ref{internalbimo} and \ref{allden} one sees that $(\sigma_1^*,\sigma_2^*,\sigma_3^*) \simeq (0.23,0.34,0.43)$ with fluctuations with standard deviations of order $0.013$. For a given configuration $(\sigma_1^*,\sigma_2^*,\sigma_3^*)$ there are two other symmetric configurations by meridional translation:
$(\sigma_2^*,\sigma_3^*,\sigma_1^*)$ and $(\sigma_3^*,\sigma_1^*,\sigma_2^*)$. However there exists also three other states that can not be recovered by meridional translation from the first one $(\sigma_2^*,\sigma_1^*,\sigma_3^*)$, $(\sigma_3^*,\sigma_2^*,\sigma_1^*)$ and $(\sigma_1^*,\sigma_3^*,\sigma_2^*)$, but can be recovered if we use a mirror symmetry. These 6 states are in correspondence with all the permutations of $\{ 1,2,3 \}$, three even permutations and three odd permutations.

\begin{figure}
\centerline{\includegraphics[width=15cm]{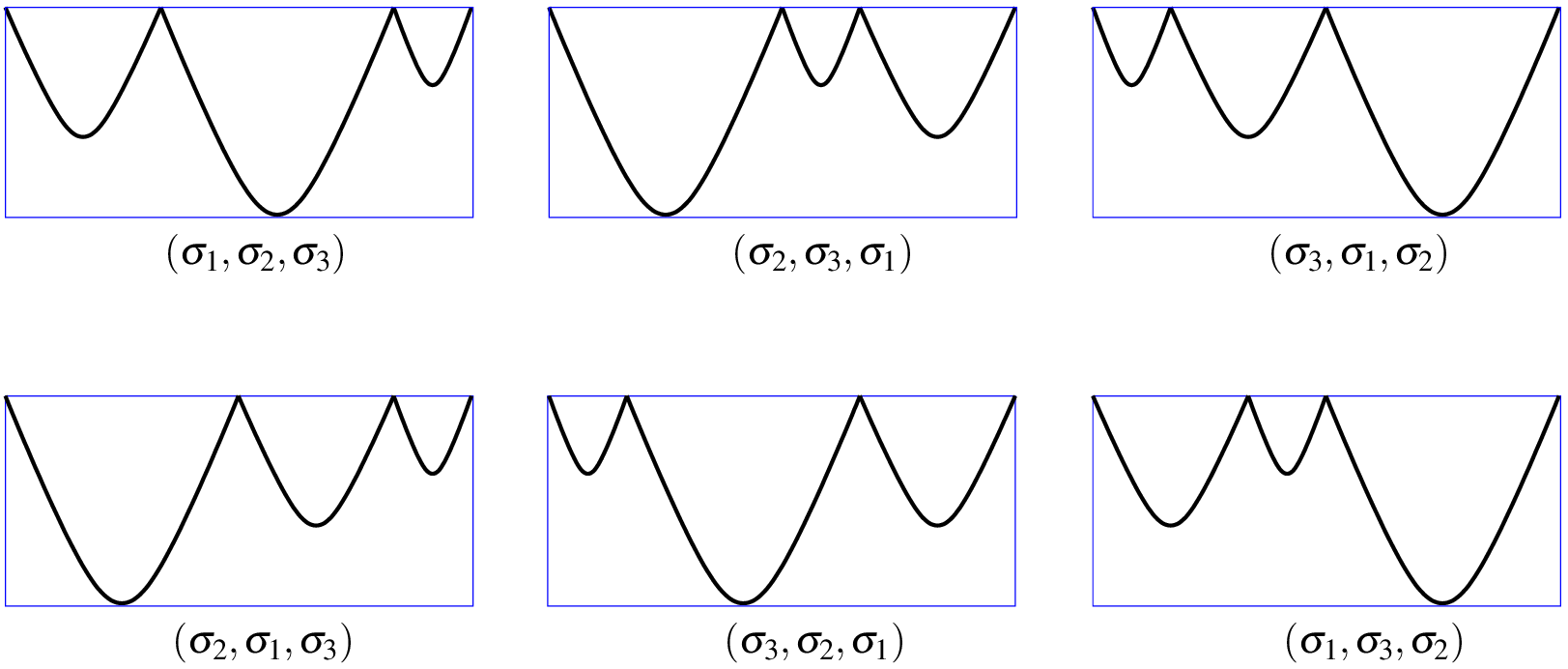}}
\caption{Sketch of the zonal velocity as a function of latitude, $U(y)$, generated by the symmetry group for the six different configurations, with three-alternating jet. We have amplified the meridional
asymmetry of the jets for better visualisation.
}
\label{perm}
\end{figure}

For clarity, Fig.\ref{perm} shows a schematic of these configurations.
The group of permutation is called ${\cal S}_3$ and possesses $3! = 6$ possible permutations: $\{ 123,231,312,213,321,132 \}$.
We now show in the next subsections a remarkable result: the symmetric group ${\cal S}_3$ is indeed realized by the natural dynamics of the system.
\subsection{Probability density in the $(\sigma_1,\sigma_2)$ plane}
\begin{figure}
\centerline{\includegraphics[width=10cm]{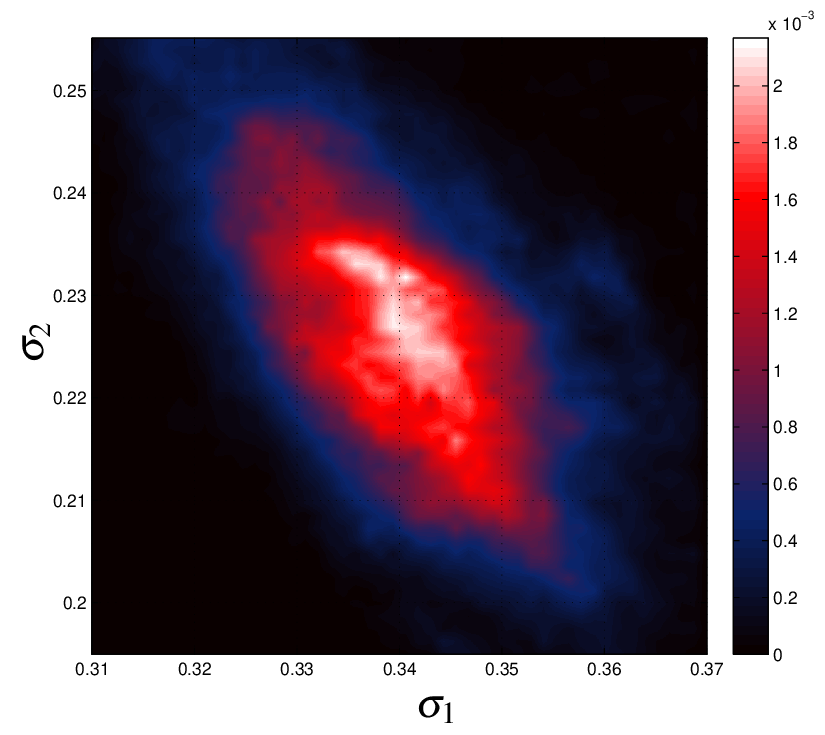}}
\caption{Probability density function of  $(\sigma_1,\sigma_2)$  for one of the states with three alternating jets ($\alpha = 10^{-3}$ and $\beta=8$).}
\label{den}
\end{figure}

We perform a  direct numerical simulation with some long time integration of equation (\ref{eq:barotropic}), and we measure the distances $\sigma_1$ and $\sigma_2$. This probability density function of $(\sigma_1,\sigma_2)$ is shown in Fig.~\ref{den}. The time averaged values obtained from this dataset gives $\langle \sigma_1 \rangle = \sigma_1^* \approx 0.340$ and $\langle \sigma_2 \rangle  = \sigma_2^* \approx 0.226$ with fluctuations for $\sigma_1$ and $\sigma_2$ being of the same order, of about $\pm 0.013$.

\begin{figure}
\centerline{\includegraphics[width=10cm]{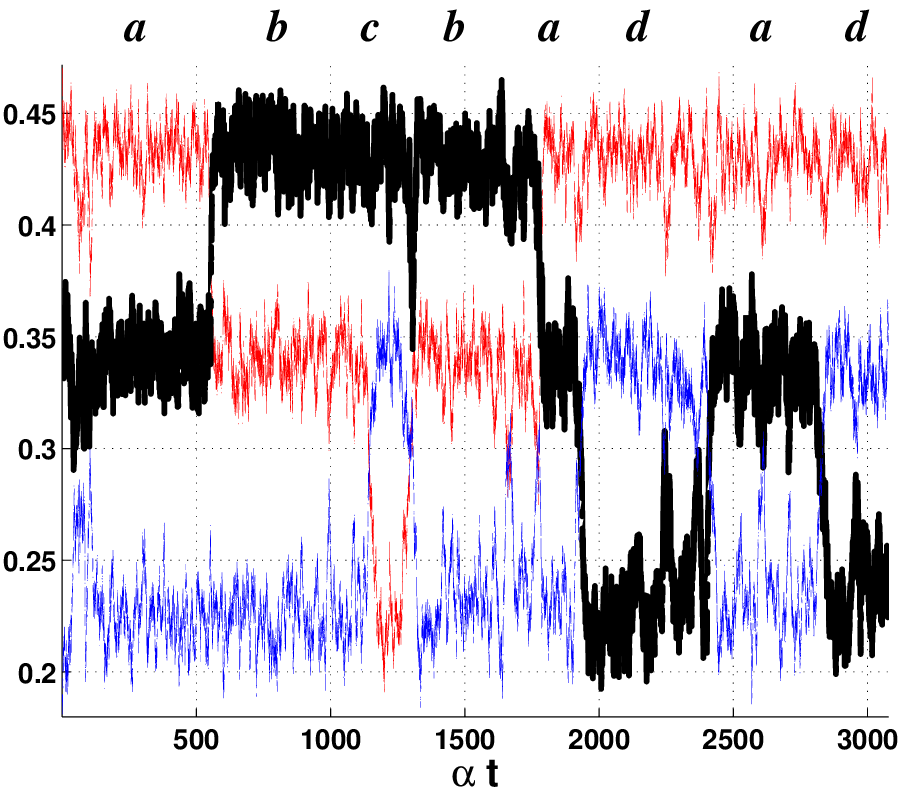}}
\caption{Timeseries of $\sigma_1$ (black curve) together with $\sigma_2$ (blue curve) and $\sigma_3$ (red curve)
with the time axis rescaled by $\alpha = 10^{-3}$.}
\label{internalbimo}
\end{figure}

Figure \ref{internalbimo} shows that when one integrates the system over a very long period of time, metastability is indeed observed between the six states.
We denote $\langle  \sigma_1 \rangle $ the value of  $\sigma_1$ averaged over a sliding time windows over a small enough period of time. It reaches three distinct
values $\langle \sigma_1 \rangle \in \{0.23,0.34,0.43\}$ corresponding to the three values  $\{\sigma_1^*,\sigma_2^*,\sigma_3^*\}$. A closer inspection of the quantities
$\sigma_1$, $\sigma_2$, shows that the system is indeed wandering around four distinct configurations denoted $a,b,c,d$ in Fig.\ref{internalbimo}. When one integrates for a sufficiently long period of time, the number of possible configurations is in fact 6, in correspondance with the 3! permutations of the symmetric group  ${\cal S}_3$.

\begin{figure}
\centerline{\includegraphics[width=10cm]{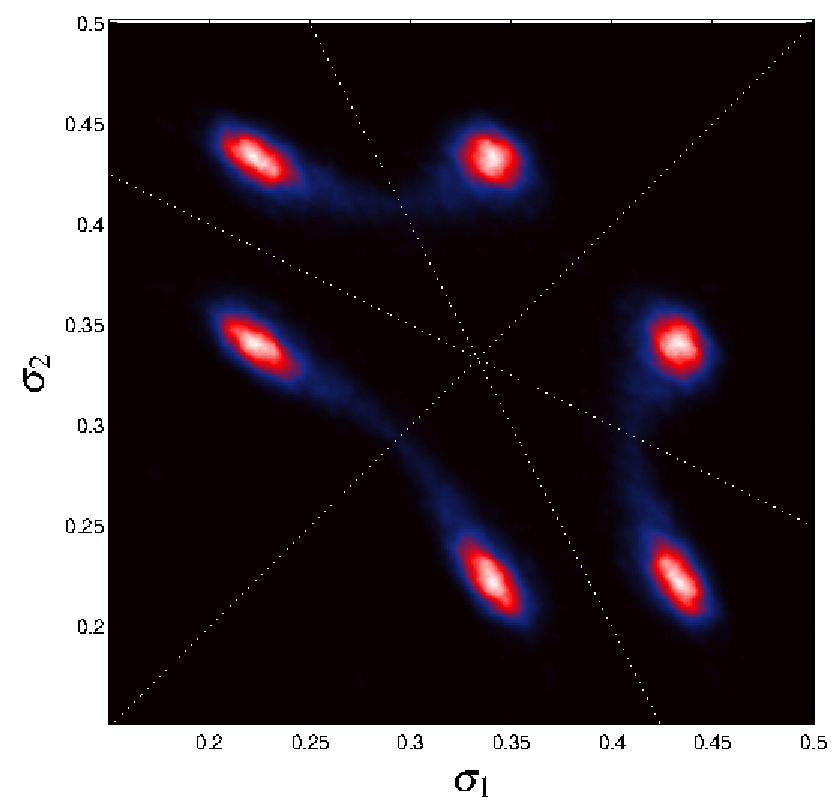}}
\caption{Probability density of $(\sigma_1,\sigma_2)$ featuring the six different states with three alternating jets. The three dotted lines correspond to states with two equally distant jets, they intersect at $(\sigma_1,\sigma_2,\sigma_3) = (\frac13,\frac13,\frac13)$.
Figure \ref{den} is a zoom of the states in the lower-left sector corresponding to configuration $a$. The blue traces inbetween the states feature reactive trajectories statistics.}
\label{allden}
\end{figure}
The probability density function of the full set is shown in Fig.\ref{allden}. One clearly sees in light blue some traces of transitions between pairs of adjacent states. As explained in the next subsection, all states are indeed connected by two type of transitions, one toward each adjacent state.
\subsection{Internal saddles}
\begin{figure}
\centerline{\includegraphics[width=12cm]{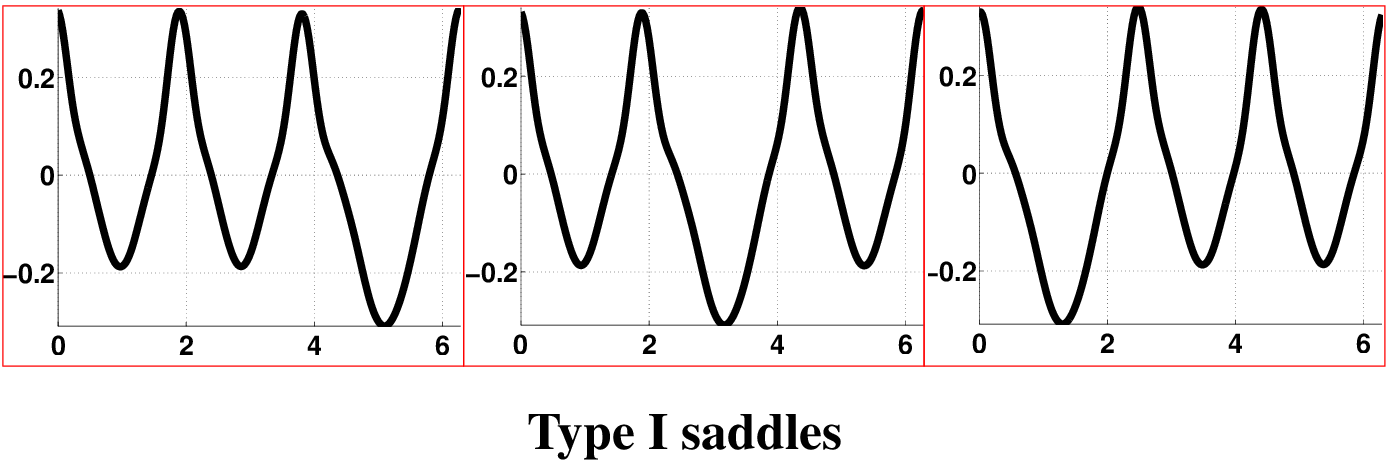}}
\caption{The three type-I saddles $(\sigma,\sigma,1-2\sigma), (\sigma,1-2\sigma,\sigma)$ and $(1-2\sigma,\sigma,\sigma)$ are located at the intersections between the dot lines and the blue traces of Fig.\ref{allden} with $\sigma \approx 0.30$. The corresponding zonally averaged zonal velocities as a function of latitude, $U(y)$, which were obtained by direct numerical simulation, are represented.
}
\label{typeI}
\end{figure}

\begin{figure}
\centerline{\includegraphics[width=12cm]{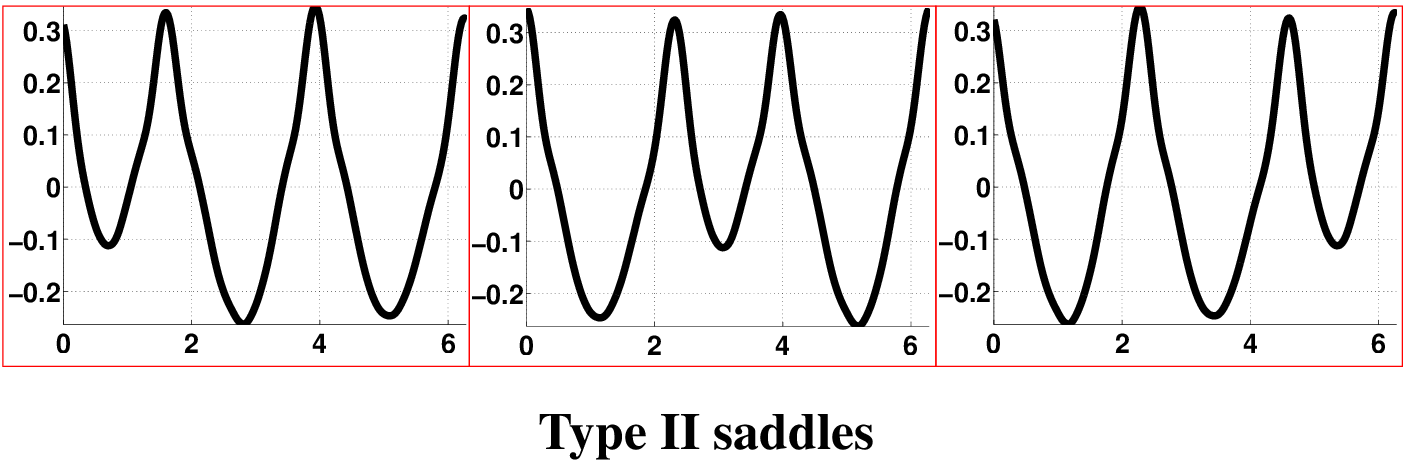}}
\caption{The three plots show the zonally averaged zonal velocities as a function of latitude, $U(y)$, for the three type-II saddles $(\sigma,\sigma,1-2\sigma), (\sigma,1-2\sigma,\sigma)$ and $(1-2\sigma,\sigma,\sigma)$ with $\sigma \approx 0.37$. They have been obtained by direct numerical simulations.}
\label{typeII}
\end{figure}

The transitions observed must involve saddles with two equally distant jets and are therefore located on the
white dot lines of Fig.\ref{allden}. Let a state be written as $(\sigma_1,\sigma_2,\sigma_3)$ and let $i,j,k \in \{ 1,2,3\}$ such that $\sigma_i < \sigma_j < \sigma_k$.
Then type-I transitions correspond to transitions where the two smallest distances have been switched so that the new state has $\sigma_j < \sigma_i <
\sigma_k$. Indeed, it can be seen in Fig.\ref{allden} by looking at the visible blue traces. This is a consequence of the two possible
nucleations of the form $(\frac14-\epsilon,\frac14+\epsilon,\frac12)$ and $(\frac14+\epsilon,\frac14-\epsilon,\frac12)$.
Type-II transitions follow the same type of rule, where instead, the two largest distances are involved. In the example above, the new state
will be such that $\sigma_i < \sigma_k < \sigma_j$. It is clear that there are no
possible transitions where the smallest and largest distances are switched, as it would require to go through the perfectly symmetric state
$(\frac13,\frac13,\frac13)$: the system prefers to visit two configurations before, through one type-I transition, and one type-II transition.
By symmetry, we conclude that there are a total of three type-I saddles and three type-II saddles. Those saddles are computed as discussed in Appendix B.c. Type-I  saddles have two equally-small distances
and are shown in Fig.\ref{typeI}. Type-II saddles have two equally-large distances and are shown in Fig.\ref{typeII}.

\subsection{Effective dynamics: summary}
\begin{figure}
\centerline{\includegraphics[width=12cm]{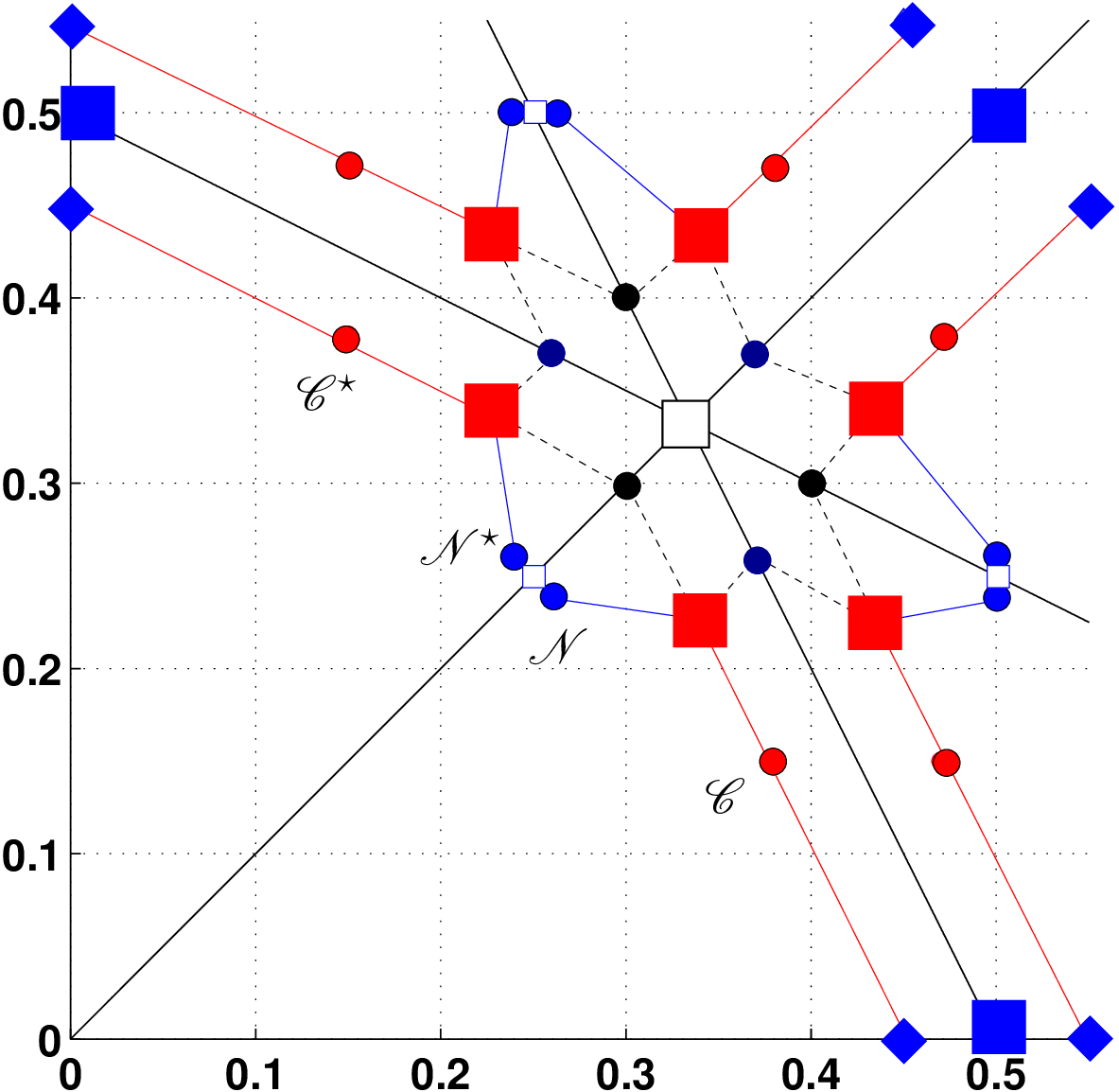}}
\caption{Effective phase space structure in the $(\sigma_1,\sigma_2)$ plane, including all ${\cal S}_3$ symmetric states. The red squares are the three-alternating jet states, the blue
squares correspond to the two alternating jet symmetric states. The blue diamonds are asymmetric two-alternating jet states obtained just after the coalescence process.
Red circles are the coalescence saddles, blue circles are the nucleation saddles, black circles are
type-I internal saddles and dark blue circles are type-II saddles. The middle square is the (unstable) three alternating jet symmetric state. The states just before the nucleations are shown as small white squares. The positions correspond to observed numerical values. The dot lines are the internal mirror
transitions responsible for metastability and the solid blue and red lines are the $2\to 3$ and $3 \to 2$ transitions, respectively.
}
\label{eff}
\end{figure}

We first mention that one can represent a two-jet state in our $(\sigma_1,\sigma_2)$ parameter plane in different ways as there is one
missing jet. We decide to choose the one which is consistent with the dynamics in the limit where one jet appears or disappears.
One must identify first the four corners $(0,0) \equiv (0,\frac12) \equiv (\frac12,0) \equiv (\frac12,\frac12)$ as a single state. During the nucleation process,
the initial perturbation before reaching the nucleation saddle is a jet of the form $(\frac14,\frac14,\frac12)$ (plus permutations)
and the nucleation saddle is of the form $(\frac14+\epsilon,\frac14-\epsilon,\frac12)$ and its five symmetric partners making
a total of six different nucleation saddles. One can measure the numerical value of $\epsilon$ from Fig.\ref{nuksad}, it gives $\epsilon \approx 0.01$.
The same rule applies for the coalescence saddles,
there are a total of six coalescence saddles. Figure \ref{eff} is a summary of all the results.

Despite its simplicity, one should remark that there is, in addition, sub-internal low-frequency variability
within a single of the six states with three alternating jets as observed in Fig.\ref{den}. This suggests the existence of additional saddles possibly within the white S shape visible in Fig.\ref{den}. We do not show the anti-instantons $2 \to 3 \to 2$ whereas the internal type-I and type-II pairs of instantons anti-instantons obviously coincide.
Note also that higher-order transitions do exist and complicate the picture, one can observe for instance examples where nucleations
and coalescences are superimposed, associated with transient four-jet states in addition with the double-nucleation scenario. These events occur with a smaller probability and are not studied here.

Our system is in a regime where at least two successive ${\cal S}_3$ symmetry-breaking bifurcations of the effective dynamics have occurred,
one related to type-I transitions and one to type-II.
Due to that, internal and complex low-frequency variability can induce rather large
fluctuations of the states with three alternating jets.  The scenario is in fact similar to the typical deterministic one, where global bifurcations
like homoclinic and heteroclinic bifurcations force the system to have large-amplitude oscillations.
One can readily anticipate that there are simpler regimes unfolding
from the perfectly symmetric state. Changing $\beta$ does not seem to change radically the internal structures shown in Fig.~\ref{eff},
except for the saddles involved in the transitions $2 \to 3 \to 2$ which move either to the states with two alternating jets or the states with three alternating jets, depending on
how $\beta$ is changed. It is plausible that, changing for instance the aspect ratio of the domain, could reveal situations where the
state $(\frac13,\frac13,\frac13)$ is an attractor. We did not explore these interesting possibilities.

\section{Large deviations and the Arrhenius law}\label{ldp}
We now address the important question of the dependency of the attractor structure and the transition rate between attractors, with respect to the parameter $\alpha$. From equation (\ref{effeq}), as a direct consequence of
Freidlin-Wentzell theory \citep{FW}, a large-deviation principle for the full system
can be inferred. The main ideas of Freidlin--Wentzell large deviation theory are sketched in Appendix A.

In particular, an Arrhenius law must be present. If we call $T$ the mean waiting time before a spontaneous transition occurs, an Arrhenius law with large deviation rate $\alpha$ is a relation $T \propto {\rm exp}\left(\frac{\Delta V}{\alpha} \right)$. $\Delta V$ can be computed from $F$ and $\sigma$ in equation (\ref{effeq}) by formula (A2) in appendix A. Moreover, as explained in section 3\ref{saddle}, the existence of $F$ and $\sigma$ and their definition from equation (\ref{eq:barotropic}) are justified in the limit $\alpha \rightarrow 0$, in the theoretical works \cite{Bouchet_Nardini_Tangarife_2013_Kinetic_JStatPhys} and \cite{bouchet2018fluctuations}, respectively.  Moreover those works explain how to compute $F$ and $\sigma$ in principle from equation (\ref{eq:barotropic}). This gives a way to possibly compute $F$, $\sigma$ and $\Delta V$ numerically from (\ref{eq:barotropic}) and (A2). However, this is a very complex numerical/theoretical program that has not been performed so far. In this section, we will follow another route and check through direct numerical simulations and the AMS algorithm the Arrhenius law.\\

A consequence of the large-deviation principle is that the effective dynamics described in Fig.~\ref{eff} is indeed valid in the limit $\alpha \to 0$,
so that reactive tubes, such as those shown in Fig.~\ref{TR}, in this case concentrate near well-defined instanton paths.


\begin{table}
\caption{Mean values of $\sigma_1^*$ and $\sigma_2^*$ for the jet attractors shown in Fig.\ref{den} as a function of $\alpha$.}\label{t1}
\begin{center}
\begin{tabular}{ccccrrcrc}
\toprule
$\alpha$ & $<\sigma_1>$ & $<\sigma_2>$ & std \\
\midrule
 $8 \cdot 10^{-4}$  & 0.342 & 0.213 & $1\cdot 10^{-2}$   \\
 $6 \cdot 10^{-4}$  & 0.342 & 0.202 & $8\cdot 10^{-3}$   \\
 $4 \cdot 10^{-4}$  & 0.340 & 0.191 & $6\cdot 10^{-3}$   \\
 $2 \cdot 10^{-4}$  & 0.334 & 0.177 & $4\cdot 10^{-3}$   \\
\bottomrule
\end{tabular}
\end{center}
\end{table}

\begin{figure}
\centerline{\includegraphics[width=12cm]{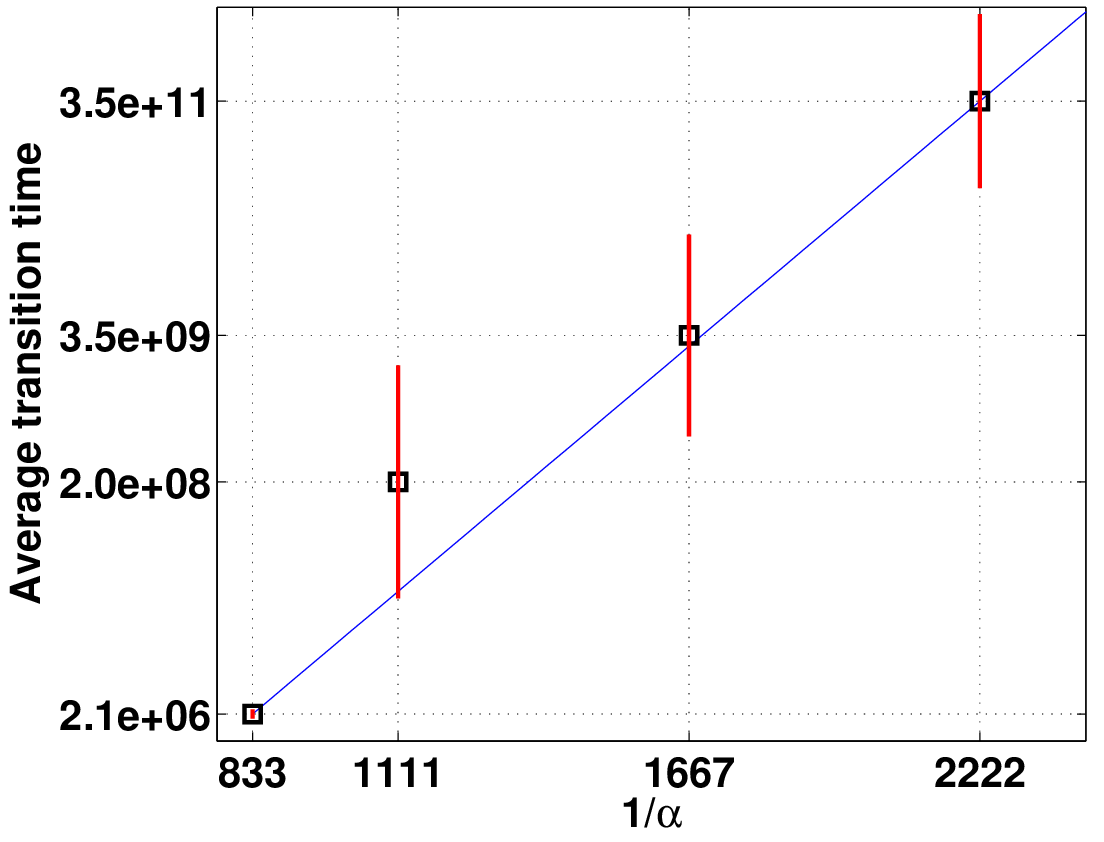}}
\caption{Logarithm of the mean first transition time $T$ as a function of $\frac{1}{\alpha}$ for the $2 \to 3$ nucleation transitions. The number of clones for the AMS algorithm
is $N=1000$, and two to three independent algorithm realizations were performed. Error bars are a crude approximation of the AMS variance based on the different realizations.
}
\label{Arr}
\end{figure}

We now discuss our last experiment aimed at checking Arrhenius law.  Using the
AMS algorithm, we compute for different values of $\alpha$, the mean waiting time to observe the $2 \to 3$ nucleation transitions.
Note that we have used a different value of $\beta=5.5$ instead of $\beta=8$. The dependency on $\beta$ is in fact irrelevant, as the result we discuss here
is mostly insensitive to $\beta$. Note also that although the probability of  $2\to 3$ transitions actually depends on $\beta$, the nucleation phenomenology remains
unchanged. The result is shown in Fig.\ref{Arr} and is consistent with the existence of an Arrhenius law of the form $T \propto {\rm exp}\left(\frac{\Delta V}{\alpha} \right)$. Such an Arrhenius law is a remarkable fact.

If one decreases $\alpha$ even further, the scaling observed in Fig.\ref{Arr} does not correspond anymore to the simple Arrhenius law (not shown). This is most probably because viscosity $\nu$ is not negligible anymore for so small values of $\alpha$. Then, the working hypothesis $\nu \ll \alpha$ in order to obtain $T \propto {\rm exp}\left(\frac{\Delta V}{\alpha} \right)$ is no more valid. To support this hypothesis, one can observe that the mean energy expected to be equal to one for negligible values of $\nu$, drops to values significantly smaller than one when one uses such small values of  $\alpha$. Using smaller value of $\nu$ to test further the Arrhenius law would require to make much more costly numerical simulation with a higher resolution, which would be extremely difficult.

It is important to understand that such scaling law cannot be revealed by direct numerical simulations as illustrated by the table \ref{t2}:
The amount of required CPU time to study the Arrhenius law with a direct numerical simulation increases exponentially as $\alpha$ decreases, whereas using the AMS the required CPU time increases only linearly (see Appendix B).

\begin{table}
\caption{CPU time (d:day,y:year) needed to obtain about $1000$ $2 \to 3$  transitions using 200 processors for both
AMS and DNS and for different values of $\alpha$.}\label{t2}
\begin{center}
\begin{tabular}{ccrccrrcrc}
\toprule
$\alpha$ & AMS & DNS \\
\midrule
 $1.20 \cdot 10^{-3}$  & 1.0 d & 15 d         \\
 $0.90 \cdot 10^{-3}$  & 1.4 d & 210 d        \\
 $0.60 \cdot 10^{-3}$  & 2.2 d & $\sim$ 51 y  \\
 $0.45 \cdot 10^{-3}$  & 3.4 d & $\sim$ 2050 y\\
\bottomrule
\end{tabular}
\end{center}
\end{table}

\section{Conclusion}\label{sconclusion}
Using the AMS rare event algorithm, we are able to obtain a complete statistical
description of the dynamics of planetary jets in a simple barotropic quasigeostrophic model.
We show that the spontaneous appearance of new jets come from noise-activated nucleations originating from the westward jets. The disappearance
of a jet is in fact the merging between two eastern jets as they become attracted to each other if they are too close. We call this event a coalescence.
By collecting thousands of these transitions, we show that they concentrate near some preferred transition paths suggesting the presence of instantons
in the limit of vanishing noise.  A computation of the mean first-passage time as a function of the Ekman dissipation $\alpha$ reveals a scaling law
clearly compatible with an Arrhenius law. This implies that the observed phenomenology is indeed instanton-driven.
As a by-product of our rare event algorithm, we are also able to compute the saddles of the effective dynamics.

We next focus on the internal dynamics of solutions having three eastward and westward jets. Surprisingly, it shows a striking example of internal
metastability. The state with three alternating jets appears to have jets not at equal distance to each other. We show that there are in fact six distinct
configurations which are governed by the symmetric group of permutation called ${\cal S}_3$.  The consequence is a complex dynamics with
low-frequency variability and large fluctuations. This behavior originates from at least two ${\cal S}_3$ symmetry-breaking bifurcations of the effective dynamics
associated with two different types of transitions depending on the distance between the jets.

We stress that these results are generic. For instance, the observed internal metastability of the three-jet states should appear as soon as there is some asymmetry in the jet positions and shapes. It comes from deeper algebraic constraints linked to the symmetries of the solutions. To this respect,
changing $\beta$ should not alter this picture but rather has an impact on the stability of these states.
As discussed before, changing $\alpha$ to even smaller values will not change this picture either.

One expects each metastable state to concentrate on a well-defined single configuration.
One may also ask whether the use of a doubly-periodic domain has an impact on these results or not?
As far as transitions between states with two and three alternating jets are concerned,
the nucleation and coalescence phenomenology is mostly local and should prevail in more general domains. The internal symmetry-breaking
dynamics of the state with three alternating jets is likely to be modified in more visible ways. However, we expect this mechanism to be robust,
since it is the relative distance between the jets which plays a role. As a basis for these assertions, we demonstrate that this phenomenology is preserved when one uses lateral meridional walls in Appendix~\ref{awall}.
As mentioned in section \ref{internald}, it is plausible that changing the aspect ratio to a small value has a more radical impact on the internal dynamics.
Effective bifurcations, either supercritical or subcritical, are potentially controlled by the domain aspect ratio. For instance, one can possibly obtain states with three symmetric alternating jets which are
 stable configurations. In fact, all the visible structures shown in this work should unfold from a single symmetric state.
It is remarkable that all the classical bifurcation concepts including normal form theory are recovered here in a statistical sense through large-deviation principles.

One consequence of this approach is the possibility to infer many results on the dynamics involving $n$ jets. It is reasonable to think that in the general
case there are $n!$ distinct configurations controlled by the symmetric group ${\cal S}_n$. The situation becomes rather involved however as saddles can be
of many more different types instead of type-I and type-II for $n=3$.

We stress that it is very difficult to obtain such results using classical tools. The rare-event AMS algorithm is performing exponentially  better than any naive Monte-Carlo methods
and is particularly fitted to study large deviations. The price to pay is the difficulty sometimes to control the algorithmic variance. It essentially depends on the
choice of a good reaction coordinate.

We have not discussed one important aspect of the dynamics,
which is called jet {\it migration}, and is observed in advanced primitive-equation models of Jupiter \citep{Williams2}. It takes the form of slow drift
of the whole jet system which can be either northward or southward. Our mechanism of metastability does not explain such behavior since the relative distance
between the jets must be the same. We do however observe jet migrations either to the north or to the south in our simple model
(see e.g. Fig.~\ref{bimodns}). The AMS algorithm can be used in such a context by simply changing the reaction coordinate and is
expected to provide powerful insight on jet migration as well.

In the future, we would like to consider more realistic models and, in particular, two-layer baroclinic models \citep{Phillips} using a small
aspect ratio. These models are rather popular for the study of the midlatitude Jovian atmosphere
\citep{Williams,HH,Panetta,Kaspi}. Interestingly, one anticipates the same transition phenomenology than the one studied here,
at least in square domains. The Hovm\"oller diagram of \citet{Panetta}, Fig.~10 for instance is very suggestive
as one observes several small coalescences at the beginning and a well-defined $3 \to 4$ nucleation.
$1 \to 2$ transitions are also studied in details in \citet{Lee}.



\section*{acknowledgments}

Eric Simonnet acknowledges support from CICADA (centre de calculs interactifs) of University Nice-Sophia Antipolis. The computation of this work were partially performed on the PSMN platform of ENS de Lyon. The research leading to these results has received funding from the European Research Council under the European Union's seventh Frame- work Programme (FP7/2007-2013 Grant Agreement No. 616811). During its last stage, this publication was supported by a Subagreement from the Johns Hopkins University with funds provided by Grant No. 663054 from Simons Foundation. Its contents are solely the responsibility of the authors and do not necessarily represent the official views of Simons Foundation or the Johns Hopkins University.




\appendix
\section{Large deviation theory and instantons}
In this appendix, we sketch large deviation theory for dynamics with small white Gaussian noise and its applications to multistability. The main results were established by physicist until the 70' and then proven rigorously by mathematicians. A classical mathematical reference is \cite{FW}. An introduction for physicists is presented in \cite{bouchet2014non} and all properties of the quasipotential, relaxation and fluctuation paths are reviewed in \cite{bouchet2020boltzmann}, chapter 3.

Let us consider the stochastic partial differential equation (\ref{effeq})  written as
$
d U = F(U)~dt + \sqrt{2\alpha} \sigma(U) dW_t,
$
where $U \equiv U(x,t),  x \in \mathbb{R}$, $U(x,0) = U_0(x)$ is an initial condition and $W_t$ is a Wiener process. We consider the It\^o convention.
We define ${\cal V} = {\cal V}(U_0,U_1,T)$, the set of trajectories starting from $U_0$ and ending at $U_1$ at time $T$. Let $S =
\sigma\sigma^T$, where $\sigma^T$ is the transpose of the operator $\sigma$. A
path integral approach \citep[initiated by][]{OM} allows to formally express the transition probability from $U_0$ at time $t=0$
to $U_1$ at time $T$ as
\begin{equation}\label{pathi}
\Pr\left(U_1,T|U_0,0\right) = \int_{\cal V} {\rm exp} \left( -\frac{{\cal A}_T[U]}{\alpha}\right)~{\cal D}[U].
\end{equation}
where ${\cal D}[U]$ is a measure over paths, and the action ${\cal A}_T$ writes
$$
{\cal A}_T = \sup_P \int_0^T \left[\left< P, \dot U \right> - {\cal H}(U,P)\right]~ dt,\ \ \  {\rm with} \ \ \ {\cal H}(U,P) = \left< P, F(U)\right>  + \left< P, S(U) P \right>.
$$
Here $\left< f , g \right> = \int dx fg$ is the canonical scalar product and  ${\cal H}$ is the large deviation Hamiltonian.

Classical references on large deviation theory (for instance \cite{FW}) justify under generic hypothesis the existence of the quasipotential $V$ such that the probability to observe a flow $U$ for the invariant measure is $\mathbb{P}(U) \underset{\alpha \rightarrow 0}{\asymp} \exp \left( V(U)/\alpha \right)$ where $\asymp$ means a logarithmic equivalence. The quasiposential $V$ solves the stationary Hamilton--Jacobi equation $ {\cal H}\left(U,\delta V/\delta U \right) = 0$, and is characterized by the viscosity solutions of this Hamilton--Jacobi equation. As explained in \cite{bouchet2020boltzmann}, section 3.3 point 7, the quasipotential $V$ is a Lyapunov function of the deterministic dynamics $d U/d t = F(U)$. A very interesting occurence of a nontrivial Lyapunov function is obtained in \cite{ManfroiYoung} for the deterministic QG instead, and with a
spatially-dependent beta effect.

In general, an explicit formula for $V$ is not available. Exceptions are for gradient dynamics $d U/d t = -a\delta V'/\delta U$, where $a=\sigma \sigma^T$, or for transverse dynamics  $d U/d t = -a\delta V'/\delta U + G$ with  $\left< G , \delta V'/\delta U \right> =0$. For those two cases $V=V'$.
Other example for the barotropic turbulence with gradient and transverse decompositions are also described in \cite{BLZ}.

We now assume that the deterministic dynamics $d U/d t = F(U)$ has discrete point-attractors. Then the quasipotential $V$ can be computed from variational problems, the computation of transition rates between bassin of attraction (see below), and the study of a Markov chain that describes transitions between attractors (see for instance \cite{FW}).

A relevant example for us is the transition between two connected bassins of attraction. Then, the mean waiting time $T$, to observe a transition between attractors $U_0$ and $U_1$, follows an Arrhenius law  $T \underset{\alpha \rightarrow 0}{\asymp} \exp \left( \Delta V/\alpha \right)$, where $ \Delta V = V(U_0)-V(U_S)$ and where $U_S$ is the lower saddle (either a point or a set), that separates the two bassins of attraction.

For this simple case, this can be directly understood from the saddle-point approximation of (\ref{pathi}) in the limit $\alpha \to 0$. The large deviation result is then
\begin{equation}
\label{DeltaU}
\Delta V = \inf_{U \in \tilde{\cal V}} {\cal A}_ \infty[U],~{\rm with}~ \tilde{\cal V} = \{ U(t), \lim_{t \to -\infty} U(t) = U_0, \lim_{t \to +\infty} U(t) = U_1 \}.
\end{equation}
The minimizers of this variational problem are called {\it instanton paths}. They are the most probable paths going from
$U_0$ to $U_1$. Generically there is a single instanton that minimizes this variational problem.

We call transitions paths, paths $U(t)$ of the stochastic dynamics which start at $U_0$ and end at $U_1$.
From the relation between this variational problem and the path integral (\ref{pathi}), one can infer that transition paths concentrate close to the instanton. Typical fluctuations of this path around the instanton are of order $\sqrt{\alpha}$.

\section{Adaptive Multilevel Splitting algorithm (AMS)}
The idea of splitting indeed took its source during the Manhattan project with J.Von Neumann (unpublished)
and was sometimes referred to as Von Neumann splitting \citep{KH,Rosen2}. These ideas were more or less forgotten until the 90s
where they start to be rather popular in chemistry, molecular dynamics and networks.
These algorithms are now better understood from a mathematical point of view and have
become an important area of research in probability \citep{PdM}.
We use here the modern adaptive version of these splitting techniques which is proposed by \citet{CerouGuya}.
It is much more robust and versatile than the classical versions used previously. The general idea is to decompose a very
small probability into a chain of products of (larger) conditional probabilities. As such, one
mimics the evolution of species by performing Darwinian selections
on the system dynamics, in a controlled (unbiased) way. These types of algorithms essential perform a large number of mutations
and selections (branching) by cloning the system dynamics.
This is the reason they are sometimes called genetic algorithms although they bear several different names like
multilevel splitting, go-with-the-winner, rare event or even large-deviation algorithms. Note that they must not be confused
with the well-known family of importance sampling techniques \citep[see][]{IIS}.
\subsection{Algorithm description}
One needs a Markovian model $(\mathbf{X}_t^{\mathbf{x}})_{t \geq 0}$ with $\mathbf{X}_0 = \mathbf{x} \in H$, where $H$ is an ad-hoc functional space. Typically,
$\mathbf{X}_t$ corresponds to a stochastic partial differential equation (PDE) in $H$.
Let us consider two arbitrary sets ${\cal A}$ and ${\cal B}$ such that ${\cal A} \cap {\cal B} = \emptyset$. The goal is to estimate the probability
$p$ to enter the set ${\cal B}$, starting from the initial condition $\mathbf{x}$, before returning to ${\cal A}$. Let
$\tau_{\cal C}(\mathbf{x}) = \inf \{ t \geq 0, \mathbf{X}_t^{\mathbf{x}} \in {\cal C} \}$. The probability $p$ translates
to $p = p(\mathbf{x}) = \Pr(\tau_{\cal B}(\mathbf{x}) < \tau_{\cal A}(\mathbf{x}))$. A {\it reactive trajectory} is a particular
realisation of this probability. We note that $p(\mathbf{x}) = 0$ if $\mathbf{x} \in {\cal A}$ and
$p(\mathbf{x}) = 1$ if $\mathbf{x} \in {\cal B}$. The function $p:\mathbf{x} \mapsto p(\mathbf{x})$ is called either the {\it
committor function} or the {\it importance function} or the {\it equilibrium potential}, in mathematics. This function can be proven to solve the backward Fokker-Planck
equation for diffusive processes \citep[e.g.,][]{VdE}. Solving this PDE is out of question in large (infinite) dimension.
One then needs to define a quantity which measures how far a trajectory is escaping from ${\cal A}$. There is no unique choice and
this will depend mostly on the question asked. Let $\Phi: H \to \mathbb{R}$ a chosen function which is called {\it reaction coordinate} or {\it
observable}. One then defines
\begin{displaymath}
Q(\mathbf{x}) = \sup_{t \in [0,\tau_{\cal A}]} \Phi(\mathbf{X}_t^{\mathbf{x}}).
\end{displaymath}
For convenience $\Phi$ is renormalized such that $\Phi({\cal A})= 0$ and $\Phi({\cal B}) = 1$ although it is not necessary.
We can now give the description of the algorithm in its simplest version.
\\\\
{\it Pseudo-code (last particle)}: Let $N$ be a fixed number, $\Phi$ given and $\mathbf{x}$ some fixed initial condition.
The quantity $\widehat{p}$ is an unbiased estimate of $p$.
\begin{itemize}
\item[]{\it Initialization}:
set the counter ${\cal K} = 0$ and $\widehat{p} = 1$. Draw $N$ i.i.d. trajectories indexed by $1 \leq i \leq N$, all starting at
$\mathbf{x}$ until they either reach ${\cal A}$ or ${\cal B}$. Compute $Q_i(\mathbf{x})$ for $1 \leq  i \leq N$.
In the following we denote $Q_i$ these quantities since their initial restarting conditions might change.
\item[]
DO WHILE $\left(\min_i Q_i \leq 1\right)$
\begin{itemize}
\item[] {\it Selection}: find $i^\star = {\rm argmin}_i Q_i$ and set $\widehat{Q} = Q_{i^\star}$.
\item[] {\it Branching}: select an index $I$ uniformaly in $\{1,\cdots,N\} \setminus \{i^\star\}$. Find $t^\star$ such that
$t^\star = \inf_t \left\{ \Phi(\{ \mathbf{X}_t^{(I)}  \geq \widehat{Q} \right\}$.
Compute the new trajectory $i^\star$ with initial condition $\mathbf{x}^\star \equiv \mathbf{X}_{t^\star}^{(I)} $
until it reaches either ${\cal A}$ or ${\cal B}$. Compute the new value $Q_{i^\star}(\mathbf{x}^\star)$.
DO ${\cal K} \leftarrow {\cal K} + 1$ and $\widehat{p} \leftarrow (1-\frac{1}{N}) \times \widehat{p}$.
\end{itemize}
END WHILE
\end{itemize}
\begin{figure}
\centerline{ \includegraphics[width=12cm]{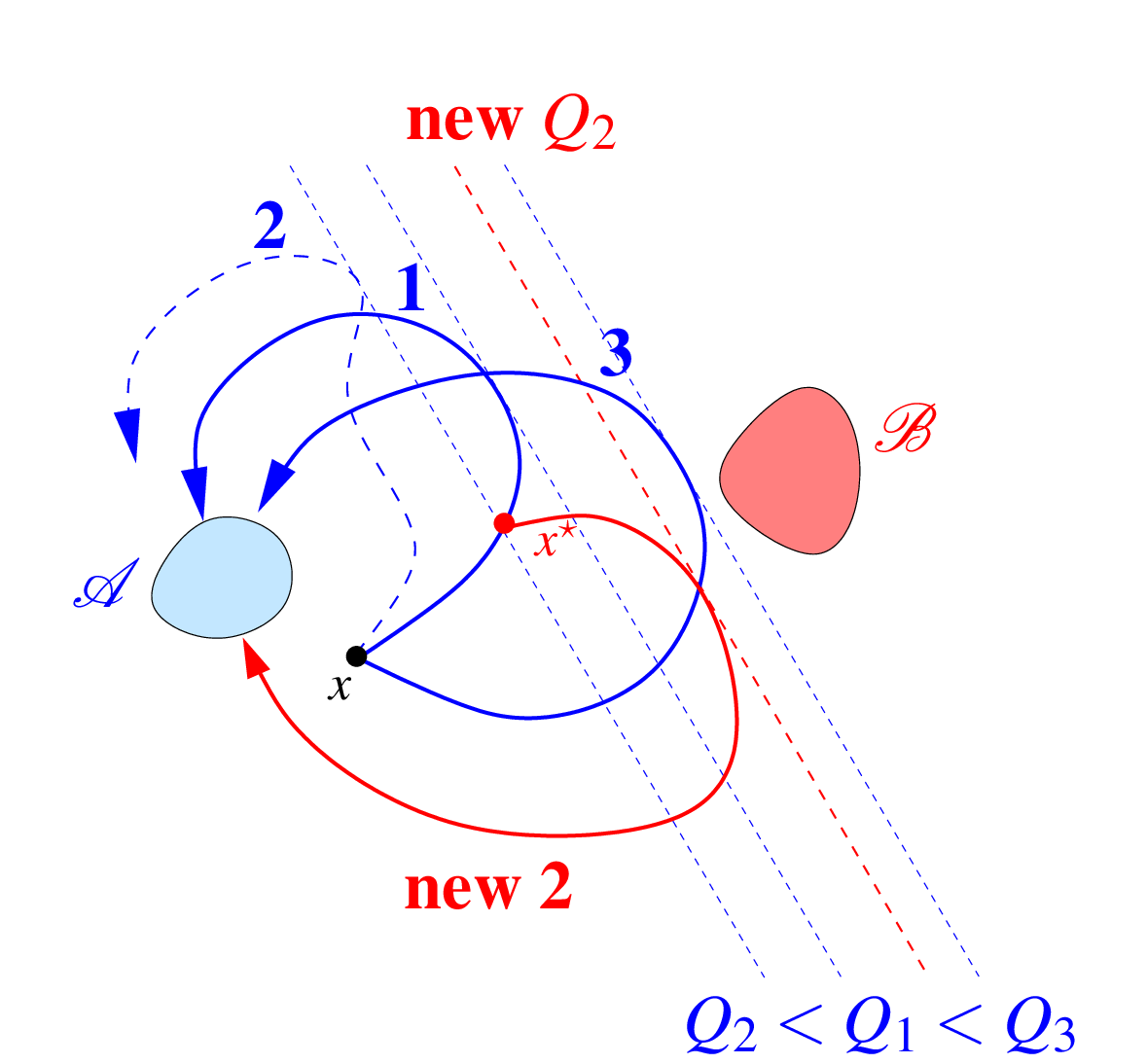}}
\caption{Sketch of the adaptive multilevel splitting algorithm (last particle algorithm) with $N=3$ trajectories.
One eliminates the trajectory (trajectory labeled 2 in the sketch) which has the smallest maximum value of $Q$ (denoted $Q_2$), and creates a new one (sketched by a red curve) which is resampled from the initial condition from one of the others trajectory chosen randomly (here trajectory 1), and when it crosses the level $Q_2$.}
\label{ams}
\end{figure}
Figure B1 gives an illustration of these algorithmic steps.
The performance of the algorithm depends crucially on the choice of $\Phi$. One can show that when $\Phi$
has the same isosurfaces than the committor $p({\bf x})$ (which is always true in 1-D), the number of
iterations ${\cal K} \sim {\rm Poisson}(-N \log p)$ \citep{last1,last2}. Central limit theorems can now be demonstrated in general situations \citep{Charles,Cerou} and typically take the form
$ \sqrt{N} \frac{p-\widehat{p}}{p} \xrightarrow[{N \to \infty}]{\cal D} {\cal N}(0,|\log p|)$. The advantage of these algorithms is the unbiased
estimate they provide. One should be careful however as a bad choice of $\Phi$ may lead to lognormal law with heavy tails and apparent bias
\citep{Joran1}.
In practice, the larger $N$ the better. A too small value of $N$ not only give results with large variance but also trigger
extinction of species by
loss of trajectory diversity. One should take advantage of algorithmic variants which eliminate $n \ll N$ trajectories at each step instead
of just one. The number of iterations ${\cal K}$ in this case scales like $\frac{N}{n} |\log p| $. Doing so in a parallel environment provide much faster results.
This is indeed the strategy used in this
paper, where we typically have $N=O(10^3)$ and $n=O(100)$ on $n$ cores with a speed-up scaling linearly like $O(n)$.
In practice, one should also run several i.i.d. algorithmic realisations with different choices of reaction coordinates
and consider a set of initial conditions $\mathbf{x}$ rather than a single initial state \citep[see e.g.,][]{lelievre}.
\subsection{Reactive coordinates}
There are many possible options to construct reasonable reaction coordinates. The first option is to consider a low-dimensional projection of the
system phase space, say on zonal Fourier modes. One can also consider EOFs or wavelets basis to better represent the solutions.
In our case, we have chosen the Fourier moduli $|q_2|,|q_3|,|q_4|$ for Fig.~\ref{TR}. One then needs to construct
a function $f(|q_2|,|q_3|,|q_4|) \in \mathbb{R}$ such that it describes
well the fluctuations of the system. Let $\mathbf{M}$ a projection of the system in the low-dimensional phase space, a natural choice is then
to consider $f(\mathbf{M}) = \frac{d_{\cal A}}{2 d_{\cal B}}$ if $d_{\cal A} \leq d_{\cal B}$ and
$f(\mathbf{M}) = 1-\frac{d_{\cal B}}{2 d_{\cal A}}$ if $d_{\cal B} \leq d_{\cal A}$, where
$d_{{\cal C}} = {\rm dist}(\mathbf{M},{\cal C})$. The function $f$ is equal to zero in the set ${\cal A}$ and one in the
set ${\cal B}$. One must be careful as the Euclidian distance may not reflect well the dynamics of the transitions. Moreover if the dimension
of the projected phase space is too low, relevant fluctuations may not be seen at all.


\subsection{Saddle detection}
Once a reactive trajectory reaches a saddle, it should relax,
with a probability close to 1, either to the set ${\cal A}$ or the set ${\cal B}$ depending whether it is "before" or "after" the saddle.
This property is at the basis of the three methods presented.
\begin{itemize}
\item[1] The simplest method requires no extra computations and is fairly accurate. Once AMS algorithm has converged, one should simply
record the states over which the last branching has occurred. The reason is that these initial conditions
all relax (by the "natural" dynamics, ie the one with probability close to 1) to the target set ${\cal B}$ by the AMS definition whereas
previous states relax to ${\cal A}$ again by definition.
\item[2] The dichotomy method is more precise than method 1, but it requires extra costly computations as one must reconstruct all AMS
trajectories \citep[see][for a similar idea]{edge,schneider2007turbulence,willis2009turbulent}.
One proceeds by dichotomy on the reactive trajectories $(\mathbf{X}_t)_{t \in [0,\tau]}$ by picking initial conditions along the trajectory.
For each initial conditions one relaxes the dynamics to decide whether it goes to ${\cal A}$ or ${\cal B}$. The dichotomy provides
a critical time, say $t_s$, and the saddle is close to $\mathbf{X}_{t_s}$.
Note that this is correct in probability only, so that one should perform several realisations, and for many AMS trajectories
In practice, one realisation is enough since the return time associated to the transition is much larger than the natural relaxation dynamics.
\item[3] This method has mainly a theoretical interest. The idea is simply to compute pairs of instantons, anti-instantons and to look for
their intersections since they must intersect at saddles and form a so-called figure-eight pattern (see Fig.~\ref{insts}).
For reversible dynamics like gradient systems, this approach is irrelevant as the geometric
support of an instanton and anti-instanton is the same. In practice, computing anti-instantons is very challenging.
\end{itemize}
%


\section{Permutations in a channel flow}\label{awall}
We investigate here the robustness of the ${\cal S}_n$ symmetric group w.r.t. the boundary conditions.
We conduct an experiment in a zonally-periodic  channel with free-slip boundary conditions, by imposing
$q = 0$ at the north and south walls and and an additional constraint $\Delta q = 0$ for the hyperviscosity.
We focus on a parameter regime having three jets ($\beta = 8$), and use the
same isotropic annular stochastic forcing than in the doubly-periodic case.
We generate $O(10)$ i.i.d. realisations starting from the same initial condition $q = 0$. We obtain
three distinct mirror pairs of solutions w.r.t. the symmetry $y \to -y$ (denoted by a bar), ie a total of 6 different
configurations.  No additional states have been observed even by increasing the number of realisations.
We define $\sigma_1$ as the distance between the first and second eastward jet and $\sigma_2$, between the
second and third eastward jet.
The results are shown in Fig.~C1 and we call $A,B,C$ the three states (right of Fig.~C1) together with their mirror counterparts $\bar A,\bar B, \bar C$ (upper part of Fig.
C1).

\begin{figure}
\centerline{\includegraphics[width=8cm]{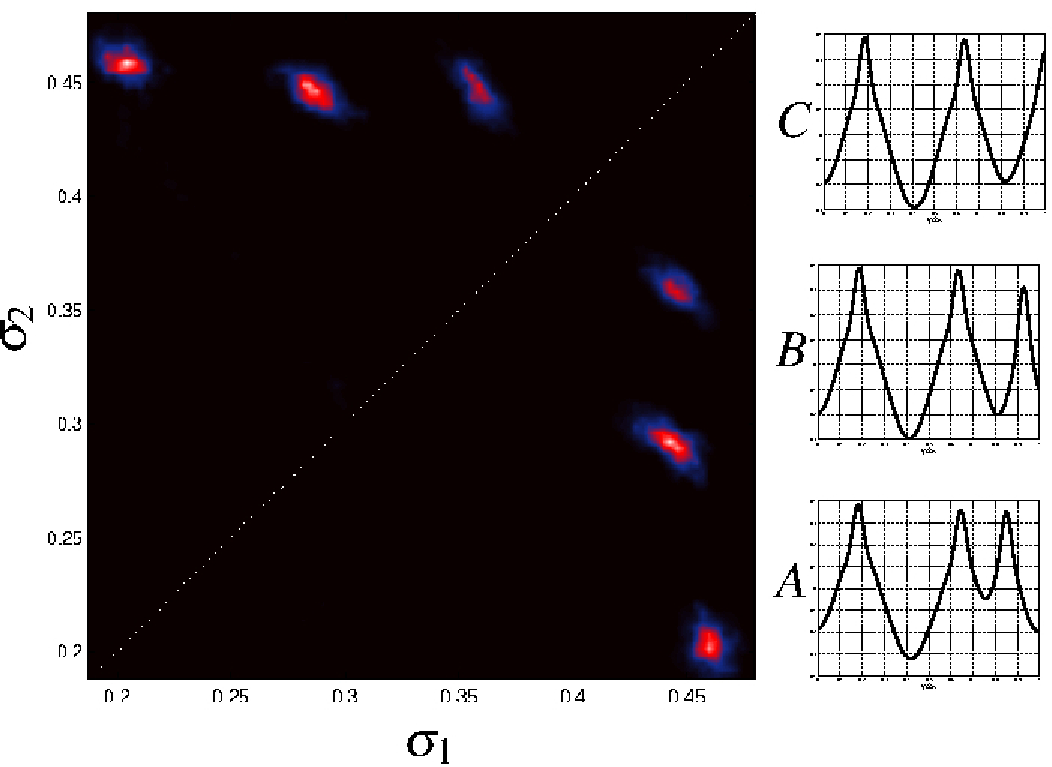}}
\caption{Probability density function $(\sigma_1,\sigma_2)$ for the 6 different three alternating jet states, in a zonally-periodic channel with free-slip boundary conditions. The computation used about $10$ independent noise realizations, each starting from rest (zero initial condition). The physical parameters are $\beta= 8$ and $\alpha = 5 \cdot 10^{-4}$. The number of grid points in each direction is $N=400$ using RK4 time scheme for an annular isotropic noise like in the doubly-periodic case at $k_f = 12$. The diagonal $(\frac13,\frac13,\frac13)$ is shown with a white dotted line.}
\label{den_chan}
\end{figure}

A close inspection of the numerical values $(\sigma_1,\sigma_2,\sigma_3)$ where $\sigma_3 = 1-\sigma_1-\sigma_2$,
shows that not all permutations are realized. Let  us look at the state $A$ with values
$(\sigma_{1,A},\sigma_{2,A},\sigma_{3,A}) \approx (0.45,0.20,0.34)$, then
$C$ is simply $(\sigma_{1,A},\sigma_{3,A},\sigma_{2,A})$, $\bar A =
(\sigma_{2,A},\sigma_{1,A},\sigma_{3,A})$ and $\bar C = (\sigma_{3,A},\sigma_{1,A},\sigma_{2,A})$,
giving a total of 4 states.
The state $B$ does not correspond to a permutation of the previous states. It has values
$(\sigma_{1,B},\sigma_{2,B},\sigma_{3,B}) \approx (0.45,0.29,0.27)$. It is almost invariant by
permuting $\sigma_2$ and $\sigma_3$. Note that the boundary conditions
impose the presence of either westward or eastward jets on the walls since
$q=0$ yields $\frac{du}{dy} = 0$, in particular we observe that
the states have either two westward jets $(A,B,\bar A,\bar B)$ or one eastward and one westward jet $(C,\bar C)$.
The existence of two eastward jets at the north and south walls seems to be forbidden for this value of $\beta$, it has never been observed.

We conclude that the symmetric group appears to be broken into two different subgroups. It suggests a
more complex general picture where the symmetries of the problem play a key role in the interaction between the jets.
It is plausible for instance that in a four-jet regime, the symmetric group ${\cal S}_3$ would emerge as a subgroup.

One of the experiment is shown in the form of an Hovm\"oller diagram in Fig.~C2.
A striking transition from $A \to \bar A$ takes place through a nucleation
inducing a 4-jet state quickly followed by a merging back to the mirror 3-jet state.
The solution goes through the perfectly marginally stable symmetry state $(\frac13,\frac13,\frac13)$ (see middle of Fig.~C2.
Transitions involving internal saddles have not been directly observed and deserve further investigations using the rare event AMS algorithm.

\begin{figure}
\centerline{\includegraphics[width=8cm]{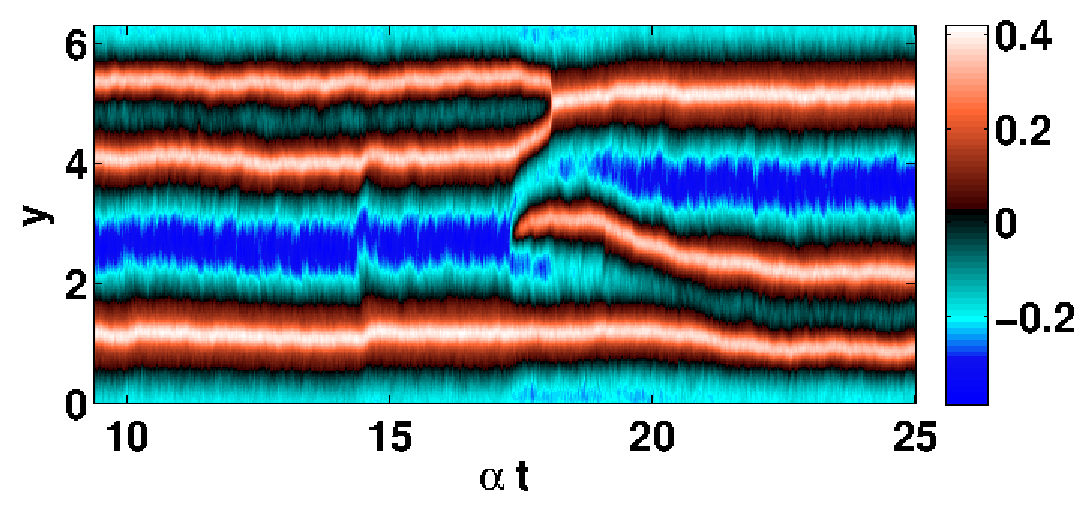}}
\caption{Hovm\"oller diagram of the zonally averaged zonal velocity in a channel flow showing a transition between
$A$ and $\bar A$.}
\label{ytran}
\end{figure}

\bibliographystyle{ametsoc2014}
\bibliography{references}



\end{document}